\documentclass[twocolumn]{aastex631}

\usepackage[T1]{fontenc} % for Icelandic Characters
\usepackage[utf8]{inputenc} % for Icelandic Characters 

\begin{document}

\title{Hipparcos, Gaia, and RVs reveal that the radio emitting F star HD 220242 has an M dwarf companion, a likely source of the radio emission\footnote{Based on observations obtained with the Hobby-Eberly Telescope (HET), which is a joint project of the University of Texas at Austin, the Pennsylvania State University, Ludwig-Maximillians-Universitaet Muenchen, and Georg-August Universitaet Goettingen. The HET is named in honor of its principal benefactors, William P. Hobby and Robert E. Eberly.}}

\author[0000-0003-1439-2781]{Megan Delamer}
\affil{Department of Astronomy \& Astrophysics, 525 Davey Laboratory, The Pennsylvania State University, University Park, PA 16802, USA}
\affil{Center for Exoplanets and Habitable Worlds, 525 Davey Laboratory, The Pennsylvania State University, University Park, PA 16802, USA}

\author[0000-0001-7409-5688]{Guðmundur Stefánsson} 
\affil{Anton Pannekoek Institute for Astronomy, University of Amsterdam, Science Park 904, 1098 XH Amsterdam, The Netherlands}

\author[0000-0001-9596-7983]{Suvrath Mahadevan}
\affil{Department of Astronomy \& Astrophysics, 525 Davey Laboratory, The Pennsylvania State University, University Park, PA 16802, USA}
\affil{Center for Exoplanets and Habitable Worlds, 525 Davey Laboratory, The Pennsylvania State University, University Park, PA 16802, USA}

\author[0000-0003-4835-0619]{Caleb I. Ca\~nas}
\altaffiliation{NASA Postdoctoral Fellow}
\affiliation{NASA Goddard Space Flight Center, 8800 Greenbelt Road, Greenbelt, MD 20771, USA}
\email{c.canas@nasa.gov}
\author[0000-0002-0872-181X]{Harish K. Vedantham}
\affil{ASTRON, Netherlands Institute for Radio Astronomy, Oude Hoogeveensedijk 4, Dwingeloo, 7991\,PD, The Netherlands}
\affil{Kapteyn Astronomical Institute, University of Groninngen,  Landleven 12, 9747AD, Groningen, The Netherlands}
\author[0000-0002-5463-9980]{Arvind F.\ Gupta}
\affil{U.S. National Science Foundation National Optical-Infrared Astronomy Research Laboratory, 950 N.\ Cherry Ave., Tucson, AZ 85719, USA}
\author[0000-0002-7167-1819]{Joseph R. Callingham}
\affil{ASTRON, Netherlands Institute for Radio Astronomy, Oude Hoogeveensedijk 4, Dwingeloo, 7991\,PD, The Netherlands}
\affil{Anton Pannekoek Institute for Astronomy, University of Amsterdam, Science Park 904, 1098 XH Amsterdam, The Netherlands}
\author[0000-0002-5093-6208]{Juan Bautista Climent Oliver}
\affil{Departament d'Astronomia i Astrof\'isica, Universitat de Val\`encia, C. Dr. Moliner 50, E-46100 Burjassot, Val\`encia, Spain}
\author[0000-0001-9662-3496]{William Cochran}
\affil{McDonald Observatory, The University of Texas, Austin Texas USA}
\affil{Center for Planetary Systems Habitability, The University of Texas, Austin Texas USA}
\author[0000-0002-3853-7327]{Rachel B. Fernandes}
\altaffiliation{President's Postdoctoral Fellow}
\affil{Department of Astronomy \& Astrophysics, 525 Davey Laboratory, The Pennsylvania State University, University Park, PA 16802, USA}
\affil{Center for Exoplanets and Habitable Worlds, 525 Davey Laboratory, The Pennsylvania State University, University Park, PA 16802, USA}
\author[0000-0003-0199-9699]{Evan Fitzmaurice}
\affil{Department of Astronomy \& Astrophysics, 525 Davey Laboratory, The Pennsylvania State University, University Park, PA 16802, USA}
\affil{Center for Exoplanets and Habitable Worlds, 525 Davey Laboratory, The Pennsylvania State University, University Park, PA 16802, USA}
\author[0000-0003-2722-1615]{Jose Carlos Guirado}
\affil{Departament d'Astronomia i Astrof\'isica, Universitat de Val\`encia, C. Dr. Moliner 50, E-46100 Burjassot, Val\`encia, Spain}
\affil{Observatori Astron\`omic, Universitat de Val\`encia, Parc Cient\'ific, C. Catedr\'atico Jos\'e Beltr\'an 2, E-46980 Paterna, Val\`encia, Spain}

\author[0009-0002-8940-7458]{Michael Hartmann}
\affil{Th{\"u}ringer Landessternwarte Tautenburg, Sternwarte 5,
D-07778 Tautenburg, Germany}
\author[0000-0002-3404-8358]{Artie P. Hatzes}
\affil{Th{\"u}ringer Landessternwarte Tautenburg, Sternwarte 5,
D-07778 Tautenburg, Germany}
\author[0009-0007-0740-0954]{Elise Koo}
\affil{Anton Pannekoek Institute for Astronomy, University of Amsterdam, Science Park 904, 1098 XH Amsterdam, The Netherlands}
\author[0000-0002-2990-7613]{Jessica E. Libby-Roberts}
\affil{Department of Astronomy \& Astrophysics, 525 Davey Laboratory, The Pennsylvania State University, University Park, PA 16802, USA}
\affil{Center for Exoplanets and Habitable Worlds, 525 Davey Laboratory, The Pennsylvania State University, University Park, PA 16802, USA}
\author[0000-0001-8720-5612]{Joe P.\ Ninan}
\affil{Department of Astronomy and Astrophysics, Tata Institute of Fundamental Research, Homi Bhabha Road, Colaba, Mumbai 400005, India}
\author[0000-0001-5654-0266]{Miguel P\'erez-Torres}
\affil{Instituto de Astrof\'isica de Andaluc\'ia (IAA-CSIC),
Glorieta de la Astronom\'ia s/n, E-18008 Granada, Spain}
\affil{School of Sciences, European University Cyprus, Diogenes street, Engomi, 1516 Nicosia, Cyprus}
\author[0000-0003-0149-9678]{Paul Robertson}
\affiliation{Department of Physics \& Astronomy, The University of California, Irvine, Irvine, CA 92697, USA}
\author[0000-0001-8127-5775]{Arpita Roy}
\affil{Astrophysics \& Space Institute, Schmidt Sciences, New York, NY 10011, USA}
\author[0000-0002-4046-987X]{Christian Schwab}
\affil{School of Mathematical and Physical Sciences, Macquarie University, Balaclava Road, North Ryde, NSW 2109, Austra}



\begin{abstract}
The detection of circularly polarized, low frequency radio emission offers the tantalizing possibility of the observation of interactions between stars and their possible substellar companions, as well as direct emission from exoplanets. Additional follow up of systems with radio emission is key to understanding the true origin of the emission, since multiple astrophysical mechanisms can plausibly lead to such signals. While nineteen M dwarfs were detected by LOFAR in circular polarization as part of the V-LoTSS survey, HD~220242 is the only F star to have a circularly polarized low frequency radio detection in the same survey. We conducted radial velocity follow up with the Habitable-zone Planet Finder and combined these observations with additional archival RVs and \textit{Hipparcos}-\textit{Gaia} proper motion accelerations to determine that HD~220242 has a stellar companion with P=16.79$\pm$0.04\,yrs and a mass of $0.619\pm0.014$\,M$_\odot$. We use Spectral Energy Distribution fitting and lack of any UV excess to rule out a co-evolved white dwarf companion and confirm that the companion is an M dwarf star. Given that F stars lack the coronal properties to produce such coherent emission, and the companion mass and lack of UV excess are consistent with an M dwarf, the radio emission is most plausibly associated with the companion.

\end{abstract}



\section{Introduction}  \label{sec:intro}
Radio emission offers the possibility to detect interactions between stars and potential substellar companions \citep{Callingham_radiosignatures_2024} enabling characterization of the space weather environment in which a planetary mass companion may evolve. Low frequency radio emission from stellar systems could also originate in the magnetospheres of exoplanets in the system and yield a direct measurement of exoplanetary magnetic field strengths \citep{Callingham_radiosignatures_2024}. Recently, state of the art low frequency radio telescopes have obtained the sensitivity necessary to probe stellar sources not just in intensity, but also in polarization. The LOw Frequency ARray \citep[LOFAR;][]{vanHaarlem_lofar_2013} has conducted a full survey of the entire northern radio sky: the LOFAR Two meter Sky Survey \citep[LoTSS;][]{Shimwell_lofar_2017}. LoTSS was aimed at tracing galaxy evolution and large scale structure. A second component of the survey looked at the circularly polarized radio emission across $\sim$27\% of the northern sky \citep[V-LoTSS;][]{Callingham_vlotss_2023}. 

Radio emission is parameterized by total intensity (Stokes I), degree of linear polarization (Stokes Q and U), and degree of circular polarization (Stokes V). V-LoTSS takes advantage of the fact that polarization can include a host of information about the emission mechanism and physical properties of the source that cannot be found in the total intensity \citep[and references therein]{wielebinski_history_2012}. Circularly polarized sources can also be more reliably associated with optical counterparts since the radio sky is significantly more sparse in Stokes V than in Stokes I \citep{callingham_blind_2019}.

While radio emitters include a wide variety of stellar and extragalactic sources, emission from extragalactic sources is often driven by the synchrotron mechanism; this mechanism does not produce a significant degree of circular polarization \citep{Legg_elliptic_1968, Saikia_polarization_1998}. A high fraction of circular polarization is expected from stellar sources, such as chromospherically active stars \citep{Slee_long_2003}, RS CVn binaries \citep{Toet_coherent_2021}, pulsars \citep{clark_polarization_1969}, star-planet interactions \citep{Saur_magnetic_2013}, and direct emission from giant substellar companions \citep{vedantham_direct_2020}. 

LOFAR is well-suited for seeking out signals that may be associated with exoplanets; the electron-cyclotron maser instability (ECMI) mechanism responsible for radio signals caused by planetary aurorae has a high frequency cutoff that is proportional to the upper limit of the magnetic field strength in the ionosphere \citep{Treumann_electron_2006}. For example, Jovian radio emission cannot be observed above $\sim$40\,MHz due to this cutoff \citep{carr_physics_1983}. While LOFAR and future arrays (e.g., Square Kilometer Array \citep[SKA][]{Schilizzi_square_2004}) offer exciting new methods for the detection and characterization of exoplanets, spectroscopic observations are needed to rule out possible astrophysical false-positive scenarios.

V-LoTSS was able to characterize the observed non-degenerate galactic sources as either M-dwarfs, chromospherically-active binaries, or sub-giant branch stars \citep{Callingham_vlotss_2023, Koo_spectroscopic_2025}. There was a single main-sequence star in the sample that was not an M-dwarf or known binary, HD~220242. It was among the most luminous stellar sources in the survey and had a very high degree of circular polarization (78$\pm$16\%).

HD~220242 is an F5 \citep{Harlan_mk_1969} star with a mass of 1.60\,M$_\odot$ \citep{Holmber_geneva_2007} located at a distance of 69.72\:pc \citep{Gaia_dr3}, making it the most distant member of the LOFAR sample on the main sequence. The star is inactive, with no known spectroscopic markers to indicate the type of coronal or chromospheric activity that could cause circularly polarized radio emission, and has an X-ray luminosity consistent with a lone main sequence F star \citep{Suchkov_ROSAT_2003}. Initial spectroscopic observations from \cite{Nordstrom_radial_1997} show no evidence of spectral lines from a secondary and lack the excess radial velocity signature expected from a close binary. 
%HD~220242 has a subsolar metallicity, which, based on the mass-metallicity trend (citation), decreases the possibility that the star has giant planetary companion that could be responsible for the observed emission. 

However, there is evidence for a long term ($>$10\,au) low mass companion in the astrometric data from \textit{Hipparcos} and \textit{Gaia} for HD~220242 \citep{Kervella_stellar_2019,kervella_stellar_2022}. Previous radial velocity (RV) analysis by \citet{Hartmann_mass_2019} suggested the companion was likely a K dwarf. This work was a summary of a long-running exoplanet survey of both F~stars and Ap~stars --- chemically peculiar A stars, so analysis on individual systems was presented briefly in cases with clear binaries. Circular and elliptical fits to the data gave minimum masses of 1.06$\pm$0.10\,M$_{\odot}$ and 0.598$\pm$0.026\,M$_{\odot}$ respectively, leading the authors to conclude a likely K~dwarf based on mass alone. Herein we present additional observations of the system obtained with the Habitable-zone Planet Finder (HPF) that demonstrate the companion has a mass of 0.619$\pm0.014$\:M$_\odot$, consistent with that of an early M dwarf. We discuss the different potential scenarios that are compatible with the available data and the radio emission observed by LOFAR.

\section{Observations} \label{sec:obs}
\subsection{Spectroscopic Data}
Our analysis makes use of recently obtained high-resolution echelle spectra from the Habitable-zone Planet Finder (HPF), as well as archival RV data from the Tautenberg Coud{\'e} Echelle Spectrograph \citep[TCES;][]{Hartmann_mass_2019} and from \citet{Nordstrom_radial_1997}.
%\subsection{Habitable-zone Planet Finder}

HPF is a fiber-fed near-infrared (NIR) precision spectrograph \citep{Mahadevan_hpf_2012, mahadevan_hpf_2014} with milli-Kelvin temperature stability \citep{Stefansson_temp_2016} on the Hobby-Eberly Telescope (HET) at McDonald Observatory \citep{ramsey_early_1998, Hill_hetdex_2021}. The HET is a fixed altitude telescope with a roving pupil design operated entirely through queue observing conducted by observatory astronomers \citep{Shetrone_queue_2007}.

We observed HD~220242 with HPF beginning on 2022 June 19 and concluding on 2023 July 23, obtaining 18 spectra over the course of 7 visits. The first five visits used 223.65\,s exposures, however the approaching sunrise led to a decreased exposure time of 63.9\,s for the sixth visit on 2023 January 26 to avoid saturating the detector. The final visit had a single 649.65\,s exposure. Spectra were binned by visit and the subsequent analysis was performed on the binned data. HPF is equipped with a Laser Frequency Comb (LFC) calibrator to track instrumental drift which can provide up to $\sim$20\,cm/s RV calibration precision in 10\,minute bins. To drift correct the HD 220242 observations, we followed the procedure described in \cite{Stefansson_subneptune_2020}, where we interpolated drift corrections from LFC drift-calibration exposures obtained regularly throughout HPF observing nights.

The 1D spectra were extracted from the 2D echellograms and corrected for bias, non-linearity, and cosmic rays following the procedures in \cite{Kaplan_algorithms_2019} and \cite{Ninan_habitable_2018}, respectively.
We extracted the RVs using a version of the SERVAL (SpEctrum Radial Velocity Analyzer) pipeline adapted for usage with HPF \citep{Zechmeister_spectrum_2018, Stefansson_subneptune_2020, stefansson_neptune_2023}.

\subsection{High-Contrast Imaging}
\begin{figure}
    \centering
    \includegraphics[width=1.0\linewidth]{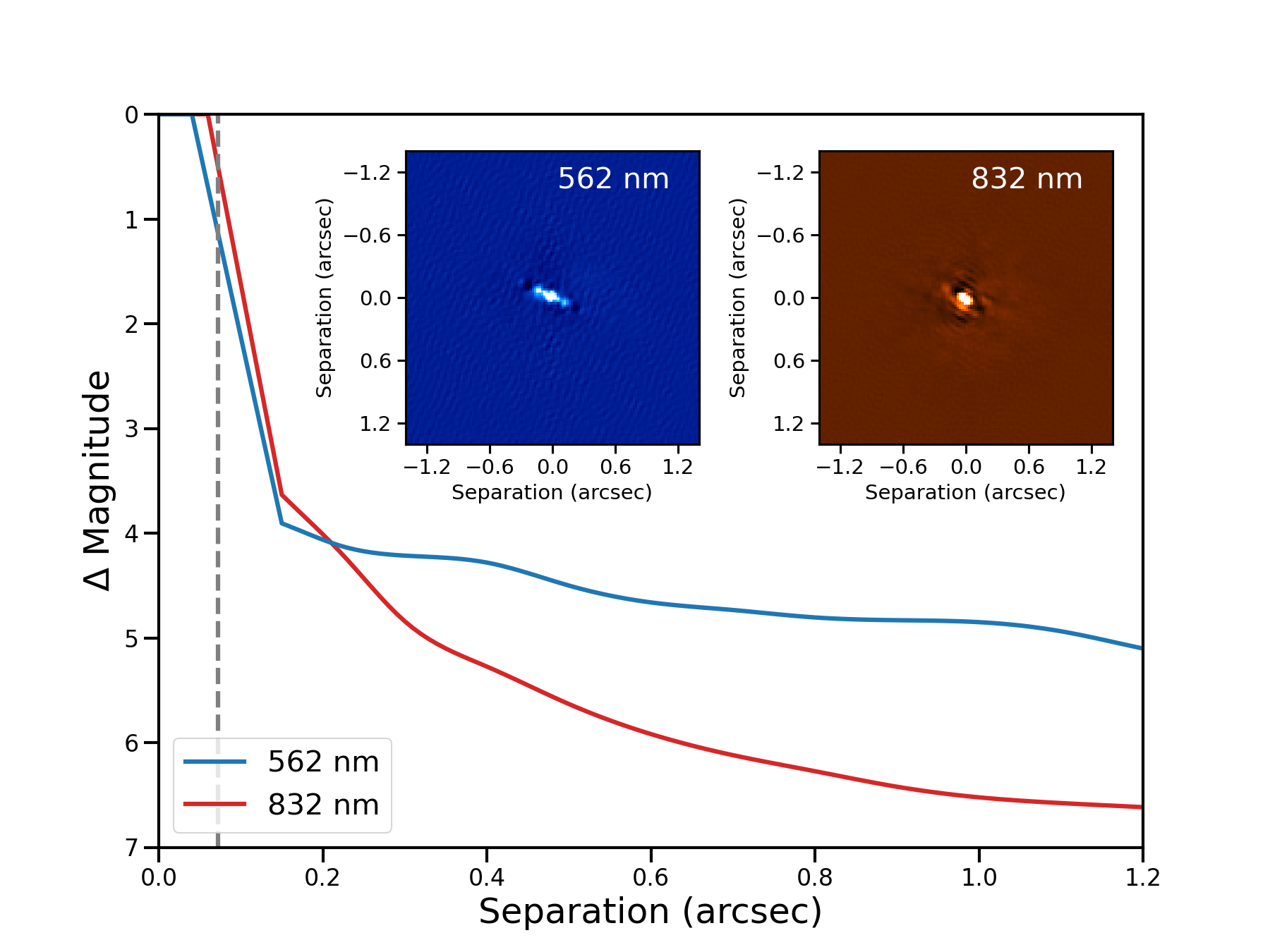}
    \caption{The 5$\sigma$ contrast curves for HD~220242 obtained from speckle imaging with NESSI in the 562 and 832\,nm filters. The insets are the NESSI speckle images centered on HD~220242A in each filter. The grey dashed line indicates the location of the companion at the time of the observations, interior to where limits can be placed. However, we find no evidence of any other bright companions.}
    \label{fig:NESSI}
\end{figure}

We observed HD~220242 on the night of 2023 February 05 with the NN-Explore Exoplanet Stellar Speckle Imager \citep[NESSI;][]{scott_nn-explore_2018} on the WIYN\footnote{The WIYN Observatory is a joint facility of the NSF's National Optical-Infrared Astronomy Research Laboratory, Indiana University, the University of Wisconsin-Madison, Pennsylvania State University, the University of Missouri, the University of California-Irvine, and Purdue University.} 3.5\,m telescope at Kitt Peak National Observatory to place constraints on the brightness of the companion. We took diffraction-limited exposures using the red camera (central wavelength of 832\,nm) at a 40\,ms cadence for 2\,minutes and reconstructed the speckle image following the methods described by \citet{Howell_speckle_2011}. We simultaneously took exposures with the blue camera (central wavelength of 562\,nm) and followed the same reconstruction procedure. We compute $5\sigma$ contrast limits as a function of separation, $\Delta\theta$, from the primary source (\autoref{fig:NESSI}). While the companion suggested by our analysis is located at too close an angular separation ($\sim0.072$") to be detected, we also find no evidence of other companions that could be responsible for the previously detected astrometric acceleration \citep{Kervella_stellar_2019, kervella_stellar_2022}.
%magnitude upper limit of 3.7\,mag in Sloan \(z^\prime\) and an upper limit of 3.9\,mag in Sloan \(r^\prime\) at the expected separation of the secondary during the observation ($\sim0.16$") \autoref{fig:NESSI}.

\section{Analysis} \label{sec:analysis}
\subsection{SED Fitting}
%\texttt{ARIADNE} is an open-source python package that utilizes Bayesian Model Averaging (BMA) to incorporate the grids of up to six different atmospheric models to obtain the temperature, log(g), [Fe/H], A$_V$, and radius of a star \citep{Vines_ariadne_2022}. The program downloads available broadband photometry (presented in \textbf{table}), fits the specified set of atmospheric models As an additional step, the package utilizes the best fitting parameters and included photometry to interpolate the mass and age of the star from MIST isochrones. The purpose of this step was twofold; we were 

\texttt{ARIADNE} is an open-source python package for spectral energy distribution (SED) fitting. It utilizes Bayesian Model Averaging (BMA) to incorporate the grids of up to six different stellar atmospheric models to obtain the effective temperature (T$_{\mathrm{eff}}$), surface gravity ($\log$ g$_\star$), metallicity ([Fe/H]), line of sight extinction (A$_V$), and radius of a star (R$_\star$) \citep{Vines_ariadne_2022}. \texttt{ARIADNE} downloads all available broadband photometry for a specified target, from the FUV band of GALEX \citep{bianchi_galex_2011} to W2 from WISE \citep{wright_wide-field_2010}, and discards any photometric points that are flagged within their catalogs. From there, the specified subset of atmospheric models are corrected for distance and each grid is fit separately. The BMA method calculates the weighted average of the estimates for temperature, $\log$ g$_\star$, [Fe/H], A$_V$, and stellar radius from each of the models, which accounts for biases inherent to individual models. As an additional step, \texttt{ARIADNE} utilizes the best fitting parameters and included photometry to interpolate the mass and age of the star from MIST isochrones  \citep{MIST}. 

\begin{figure}[h!]
    \centering
    \includegraphics[width=0.45\textwidth]{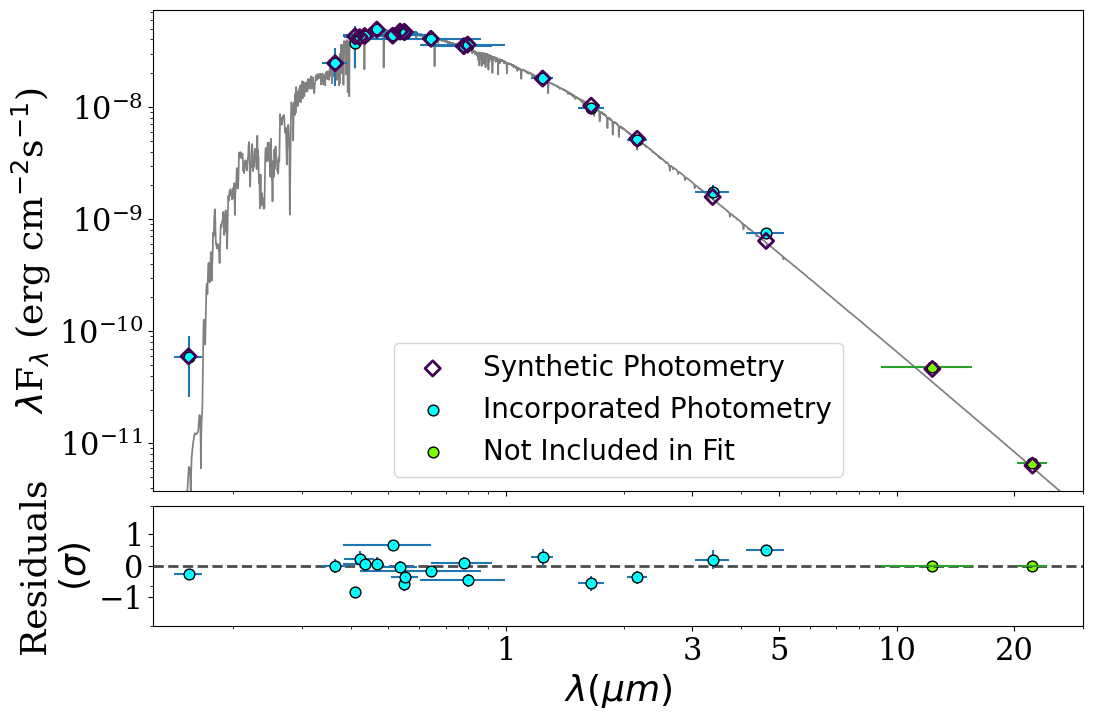}
    \caption{The available HD~220242 photometry (aqua and green circles) plotted alongside the synthetic photometry derived from convolving the best-fitting atmospheric model with the bandpass of each instrument, which are plotted as purple diamonds. Photometric points in green were not used in the fit, as \texttt{ARIADNE} uses photometric points redder than W2 solely to visualize potential IR excess; they are included only for completeness. The BTSettl-AGSS2009 model (grey line) with characteristics most similar to those found in the \texttt{ARIADNE} fit is included to guide the eye, but is not expected to be identical to the photometry. Below: The residuals between the synthetic and observed photometry.}
    \label{fig:sedfit}
\end{figure}

Our purpose in performing this fit is twofold; first, we needed to determine the characteristics of the primary to inform our orbit fitting and second, we looked for signs of either blue or red excess that could indicate the nature of the companion. For our \texttt{ARIADNE} fit, we used the atmospheric models of PHOENIXv2 \citep{Husser_new_2013}, BTSettl-AGSS2009 \citep{allard_models_2012}, \citet{kurucz_1993}, and \citet{Castelli_2003}; the other available models are identical to BTSettl-AGSS2009 above 4000\,K and thus were not used. These model grids are convolved with the filter response function for each retrieved catalog photometric data point to create synthetic photometry. The SED is then modeled by interpolating across the model grids to generate these synthetic points and the comparison is performed between the synthetic points and the retrieved catalog data. We set all priors to the default values, except T$_{\mathrm{eff}}$, which we set using the range from 6600-7000\,K. The parameters obtained from this fit are located in \autoref{tab:stellarparam} and the photometry is plotted alongside the best-fitting BTSettl model in \autoref{fig:sedfit}. 

The results are consistent with earlier work \citep{Casagrande_new_2011} and there is no indication of red excess. While the GALEX FUV photometry (highlighted in \autoref{fig:sedfit}) seems high relative to the plotted atmospheric model, the synthetic photometry (purple diamond) that is produced by convolving the model with the GALEX bandpass is consistent with the observed value at $<$1$\sigma$ and therefore there is no blue excess either. We conclude that any stellar companion in the system must have a very low flux in comparison to the F star primary.

\begin{deluxetable*}{lccc}
    \tablecaption{Summary of stellar and orbital parameters for the HD~220242 binary system. \label{tab:stellarparam}}
    \tablehead{\colhead{~~~Parameter}&  \colhead{Description}&
    \colhead{Value}&
    \colhead{Reference}}
    \startdata
\multicolumn{4}{l}{\hspace{-0.2cm} Main identifiers:}  \\
~~~TIC & \textit{TESS} Input Catalogue  & 28393683 & Stassun \\
~~~2MASS & \(\cdots\) & J23215825+2636321 & 2MASS  \\
~~~Gaia DR3 & \(\cdots\) & 2844641816070874240 & Gaia DR3\\
\multicolumn{4}{l}{\hspace{-0.2cm} Equatorial Coordinates and Proper Motion:} \\
~~~$\alpha_{\mathrm{J2000}}$ &  Right Ascension (RA) & 350.492$\pm0.031$ & Gaia DR3\\
~~~$\delta_{\mathrm{J2000}}$ &  Declination (Dec) & 26.6089$\pm0.038$ & Gaia DR3\\
~~~$\mu_{\alpha}$ &  Proper motion (RA, mas/yr) &  -88.778$\pm0.046$ & Gaia DR3\\
~~~$\mu_{\delta}$ &  Proper motion (Dec, mas/yr) & -75.869$\pm0.053$& Gaia DR3 \\
~~~$\varpi$ & Parallax & 14.4243$\pm0.0598$ & Gaia DR3\\
~~~$d$ &  Distance in pc  & $69.3_{-0.3}^{+0.3}$ & Bailer-Jones \\
\multicolumn{4}{l}{\hspace{-0.2cm} Optical and near-infrared magnitudes:}  \\
~~~$B$ & Johnson B mag & $6.999 \pm 0.004$ & APASS\\
~~~$V$ & Johnson V mag & $6.593 \pm 0.007$ & APASS\\
~~~$T$  & \textit{TESS} magnitude & $6.1935\pm0.0060$  & Stassun \\
~~~$J$ & $J$ mag & $5.813 \pm 0.20$ & 2MASS\\
~~~$H$ & $H$ mag & $5.714 \pm 0.027$ & 2MASS\\
~~~$K_s$ & $K_s$ mag & $5.662 \pm 0.020$ & 2MASS\\
~~~$W1$ &  WISE1 mag & $5.507 \pm 0.175$ & WISE\\
~~~$W2$ &  WISE2 mag & $5.435 \pm 0.062$ & WISE\\
~~~$W3$ &  WISE3 mag & $5.580 \pm 0.014$ & WISE\\
~~~$W4$ &  WISE4 mag & $5.545 \pm 0.038$ & WISE\\
\multicolumn{4}{l}{\hspace{-0.2cm} Soft X-ray Luminosity:} \\
~~~0.1-2.4\,keV & log($L_X$) & $29.08$ & ROSAT\\
\multicolumn{4}{l}{\hspace{-0.2cm} Primary Stellar Parameters from \texttt{ARIADNE}:}\\
~~~$T_{\mathrm{eff}}$ &  Effective temperature in K & $6840\pm30$ & This work\\
~~~$\mathrm{[Fe/H]}$ & Metallicity in dex & $-0.04\pm0.02$ & This work \\
~~~$M_\star$ &  Mass in $M_{\odot}$ & $1.56\pm0.07$ & This work\\
~~~$R_\star$ &  Radius in $R_{\odot}$ & $2.19\pm0.04$ & This work\\
%\multicolumn{4}{l}{\hspace{-0.2cm} Other Stellar Parameters:}           \\
 ~~~$\log$ g$_\star$ &  Surface gravity in cgs units & $4.1\pm0.2$ & This work \\
~~~$L_\star$ &  Luminosity in $L_{\odot}$ & $9.2\pm0.4$ & This work\\
~~~Age & Age in Gyrs & $1.7\pm0.2$ & This work\\
~~~$A_v$ & Visual extinction in mag & $0.07\pm0.02$ & This work\\
\multicolumn{4}{l}{\hspace{-0.2cm} Orbital Parameters:}\\
~~~K & RV Semi-Amplitude (km/s) & $4.40\pm0.06$ & This work\\
~~~T$_\mathrm{p}$ & Time of Periastron &$2459904\pm17$ & This work\\
~~~e & Eccentricity &$0.364\pm0.004$ & This work\\
~~~$\omega$& Argument of Periastron in Radians &$5.76\pm0.03$ & This work\\
~~~P & Period in Years &$16.80\pm0.04$ & This work\\
~~~i & Inclination ($^o$) &$106.6\pm0.8$ & This work\\
\multicolumn{4}{l}{\hspace{-0.2cm} Derived Parameters:}\\
~~~a & Semi-major Axis in au &$8.48\pm0.07$ & This work\\
~~~M$_s$ & Secondary Mass in M$_\odot$ &$0.619\pm0.014$ & This work\\ 
    \enddata
    \tablenotetext{}{References are: Stassun \citep{stassun_tess_2018}, 2MASS \citep{cutri_2mass_2003}, Gaia DR3 \citep{gaia_collaboration_gaia_2022}, Bailer-Jones \citep{bailer-jones_estimating_2021}, APASS \citep{Henden_apass_2014}, WISE \citep{wright_wide-field_2010}, ROSAT \citep{Huensch_rosat_1998}}

\end{deluxetable*}

\begin{figure*}[htb]
    \centering
    \includegraphics[width=0.9\textwidth]{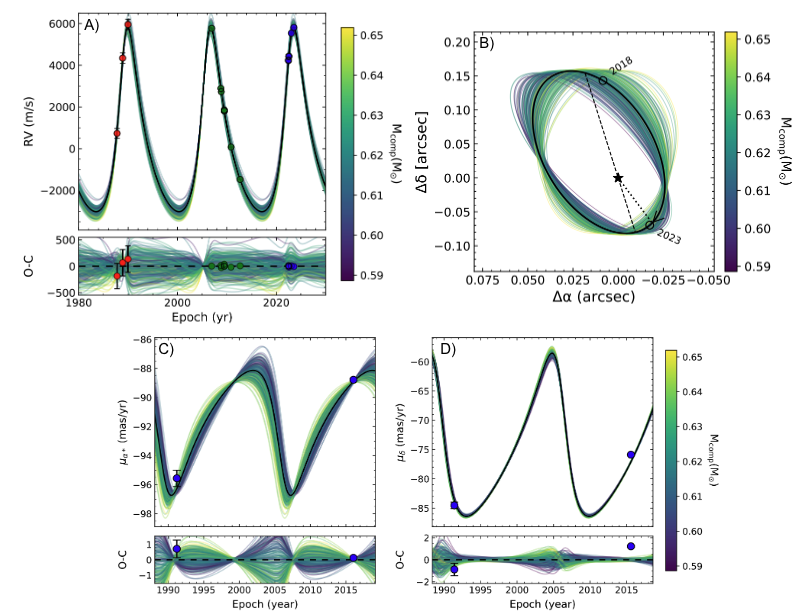}
    \caption{A)Timeseries of RV observations of HD~220242 with archival data in red \citep{Nordstrom_radial_1997} and green \citep{Hartmann_mass_2019}. The HPF RV data is in blue and the best fitting model from the \texttt{orvara} analysis is included in black. B) The projected astrometric orbit of HD~220242B, including the time stamps for the LOFAR (2018) and NESSI (2023) observations. C) The proper motion of HD~220242 in right ascension, where the two blue points indicate the \textit{Hipparcos} and \textit{Gaia} measurements. D) The proper motion of HD~220242 in declination, where the two blue points indicate the \textit{Hipparcos} and \textit{Gaia} measurements. In panels A, C, and D, the lower insets show the difference between observed and calculated points for either RVs (panel A) or astrometry (panels C and D). For all panels, the thick black line is the best fitting orbit from jointly modeling the RVs, the HGCA inputs, and the Hipparcos intermediate astrometric data and the other lines are 100 random draws from the posterior distribution, colored by the mass of the secondary. The RVs are available as data behind the figure.}
    \label{fig:pmorbit}
\end{figure*}

\subsection{Orbit Fitting with Orvara}
The Hipparcos-Gaia Catalog of Accelerations \citep[HGCA;][]{brandt_HGCA_2018} is a list of stars with known proper motion anomalies: changes in acceleration calculated based on the difference in proper motions measured by the Hipparcos and Gaia satellites. These calculated accelerations are intended to be used in searches for substellar and dark companions, not for binary stars as light from the secondary can complicate the interpretation of Hipparcos and Gaia data \citep{brandt_HGCA_2018}. However, we do not detect light from the secondary in the HPF spectra despite the small separation (see \autoref{SB2}), so we assume the measured proper motions reflect only the movements of the primary and therefore can be used in our analysis. As a check on this assumption, once we determined the companion mass and found it to be within the bounds of late K/early M, we used BTSettl-AGSS2009 model spectra \citep{allard_models_2012} for solar metallicity stars with effective temperatures between 3500-4500K to determine the expected flux ratio between the companion and the primary, finding an upper limit on contamination of 1.6\% in the broad band Hipparcos filter. This value is below the level at which significant error might be seen in the data.

We began by modeling the RV timeseries with \texttt{RADVEL}, which fits Keplerian orbits to the data to determine orbital parameters \citep{Fulton_radvel_2019}. In addition to the orbital parameters, we fit a linear trend and RV offsets for each instrument to the data to account for long-term changes from both astrophysical and instrumental causes. This initial fit allowed us to compare our results to earlier RV-only work \citep{Hartmann_mass_2019}, with which we agreed at the $<1\sigma$ threshold for all orbital parameters (orbital period, RV semi-amplitude, and eccentricity). We then used these results to provide a reasonable set of priors for the later joint fit. We assumed this massive distant companion would be the dominant astrometric signature in any Hipparcos or Gaia data due to the lack of any bright companions in the NESSI data.

\texttt{orvara} is an open-source Python package capable of combining proper motion anomaly measurements from HGCA, Hipparcos intermediate astrometric data (IAD), relative astrometry, and radial velocity data to determine orbital characteristics and estimate the dynamical masses of companions \citep{brandt_orvara_2021}. For our fit, we chose to use proper motion anomalies, Hipparcos IAD, and radial velocities. To test the Hipparcos IAD inclusion, we ran fits both with and without including this data, but the only difference was a slight increase in the errorbars when it was removed. The likelihood of \texttt{orvara} is sampled using \texttt{ptemcee} \citep{vousden_ptemcee_2016}, a parallel-tempering MCMC sampler based on \texttt{emcee} \citep{foreman-mackey_emceee_2013}. We set a prior on the primary mass using our \texttt{ARIADNE} analysis (1.55$\pm0.05$\:M$_\odot$), and set additional priors for the companion and orbit using the results from RV-only orbital modeling (orbital period, RV semi-amplitude, and eccentricity). We derive a dynamical mass of 0.619$\pm$0.014\:M$_\odot$ for the secondary, a semi-major axis of 8.48$\pm0.07$\:AU, and an eccentricity of 0.365$\pm0.006$. The RV orbit, astrometric orbit, and fit to the Hipparcos-Gaia proper motions are in \autoref{fig:pmorbit} and the complete results of the \texttt{orvara} fit are in \autoref{tab:stellarparam}.

\subsection{Spectroscopic Binary Analysis} \label{SB2}
When stellar systems are observed as double-lined spectroscopic binaries, the dynamical mass ratio, $q=M_2/M_1$, can be determined directly from the velocities \citep{mazeh_infrared_2002}, and an estimate of the primary mass from an alternate source yields the mass of the secondary. While detection of both sets of lines is nearly impossible in the optical for systems with $q$ significantly below one, the flux ratio between the two components increases as observations go towards longer wavelengths; high resolution IR spectra are better suited for searches where the companion is much less massive than the primary. Previous work has successfully detected double-lined spectroscopic binaries with mass ratios as low as 0.1 \citep{bender_detection_2008, bender_sdss_2012}, though they used spectra centered further into the IR than what we obtained from HPF.

We attempted to simultaneously measure the radial velocity of each of the two components of HD~220242 using \texttt{TODCOR}, a TwO-Dimensional CORrelation algorithm \citep{Mazeh_todcor_1994}. The algorithm calculates the correlation of an observed spectrum against a combination of two templates, using all possible radial velocity shifts. We used a selection of primary templates with $T_{\mathrm{eff}}$ from 6500-6800\,K and secondary templates with $T_{\mathrm{eff}}$ from 3700-4500\,K. We attempted \texttt{TODCOR} using both PHOENIXv2 \citep{Husser_new_2013} model spectra for the templates and using observed stars from the HPF library, but we were unsuccessful in detecting the secondary.

\section{Discussion} \label{sec:disc}
We calculate a mass of 1.55$\pm$0.05\:M$_\odot$ for the primary using isochrone analysis within \texttt{ARIADNE}, which is consistent with the literature value of 1.60$\pm0.05$\:M$_\odot$ \citep{Holmber_geneva_2007}. Our analysis also gives an age of 1.7$\pm0.2$\:Gyr for the system, placing it within the expected main sequence lifetime for a star of this mass. The $\log$ g$_\star$ derived from the SED fit is also consistent with that of a main sequence star. This consistency with the main sequence strongly biases us against the possibility of the circularly polarized emission coming from HD~220242A. While rapidly rotating sub-giants such as FK Comae Berenices variables were detected in the Stokes V survey \citep{Callingham_vlotss_2023} and are expected to have circularly polarized radio emission, main sequence F-type stars typically have magnetic field strengths of a few Gauss \citep{Seach_magnetic_2020,Seach_magnetic_2022} that are too low to generate cyclotron maser emission at 144\:MHz.

HD~220242B is not in a face-on orbit, but is slightly inclined (i=106.6$\pm$0.8$^o$) and has a mass of 0.619$\pm$0.014\:M$_\odot$, which is consistent with the minimum mass for the companion (0.598$\pm$0.026) in \citet{Hartmann_mass_2019}. This mass at the derived semi-major axis of 8.4\:AU is also in agreement with previous predictions made only from the tangential acceleration between the Hipparcos and \textit{Gaia} mean positions \citep{Kervella_stellar_2019,kervella_stellar_2022}, which indicated a companion at a minimum separation of 3\:AU.

\subsection{Possible Sources of the Radio Emission} Having determined the mass of the secondary and the orbital characteristics of the binary, we now present possible explanations for the highly polarized low frequency radio emission detected by LOFAR \citep{Callingham_vlotss_2023}. We begin by making the assumption that the emission is coming from this system and not from a background source. While it is possible for a faint radio-bright galaxy such as the ones presented by \citet{Best_LOFAR_deepfields_2023} to be concealed behind the system, extragalactic sources tend to have very low fractions of circularly polarized emission \citep[$\sim1$\%;][]{Weiler_catalog_1983,Rayner_radiocircular_2000,Agudo_polarized_2022}. The only identified extragalactic signal in the V-LoTSS sample was associated with an AGN \citep{Callingham_vlotss_2023} known to host a pc-scale radio jet \citep{Petrov_widefield_2021}; the signal from this object also had a substantially lower circularly polarized fraction ($1.4\pm0.3$\%) than HD~220242 ($78\pm16$\%).

\subsubsection{Main sequence M dwarf companion}
The secondary mass derived in our analysis is consistent with an early M dwarf. Besides HD~220242, all of the isolated main-sequence stars in the V-LoTSS sample were identified as M dwarfs, ranging in spectral type from M1.5 to M6.0. For each member of this sample, the exact mechanism that has been linked to the emission is dependent on the coronal and chromospheric activity levels as well as the rotation period of the star \citep{Callingham_population_2021,Callingham_vlotss_2023}. 

The high degree of circular polarization and high brightness temperatures in both the isolated M dwarf sample and the HD~220242 system exclude incoherent mechanisms like gyrosynchrotron emission \citep{dulk_radio_1985}, leaving plasma emission or ECMI as the most likely explanations. Plasma emission in the stellar corona is caused by the injection of impulsively heated plasma into colder plasma (e.g. stellar flares and coronal mass ejections); the non-thermal radio emission from this occurs at the fundamental plasma frequency and its second harmonic \citep{dulk_radio_1985}. This mechanism is more likely to drive the radio emission in chromospherically active M dwarfs; these stars would also be expected to have high soft X-ray luminosities. Studies of optical flares in the TESS lightcurves of the LOFAR M dwarf sample reveal an order of magnitude higher flare rate for chromospherically active members of the sample over the quiescent members \citep{pope_tess_2021}, consistent with the underlying cause of plasma emission. 

The ECMI mechanism has multiple possible drivers. The first is the breakdown of co-rotation between the magnetic field and the plasma, which can only produce the observed luminosities for stellar rotation periods $\leq$2\:days \citep{Nichols_magnetosphere_2011, Nichols_origin_2012, Callingham_population_2021}. This is a possible explanation for the emission from fast-rotating systems without chromospheric activity signatures, but previous radio bursts associated with this mechanism have been shorter in duration \citep[$\leq$20\:min;][]{Zic_askap_2019}. HD~220242 was directly observed in only one of LOFAR's eight hour observing blocks. The radio flux was consistent with a flat constant radio emission across the entire observation with no obvious bursts. ECMI radiation from M dwarfs could also be driven by a planet orbiting within the Alfv\'en surface; an explanation that has been offered for the slowly rotating, quiescent members of the LOFAR M-dwarf sample \citep{Callingham_population_2021}. Finally, the radiation could be auroral emission due to stellar wind-magnetosphere interactions between the M dwarf and a massive substellar companion \citep{Farrell_possibility_1999}.

\begin{figure*}[htb]
    \centering
    \includegraphics[width=0.9\textwidth]{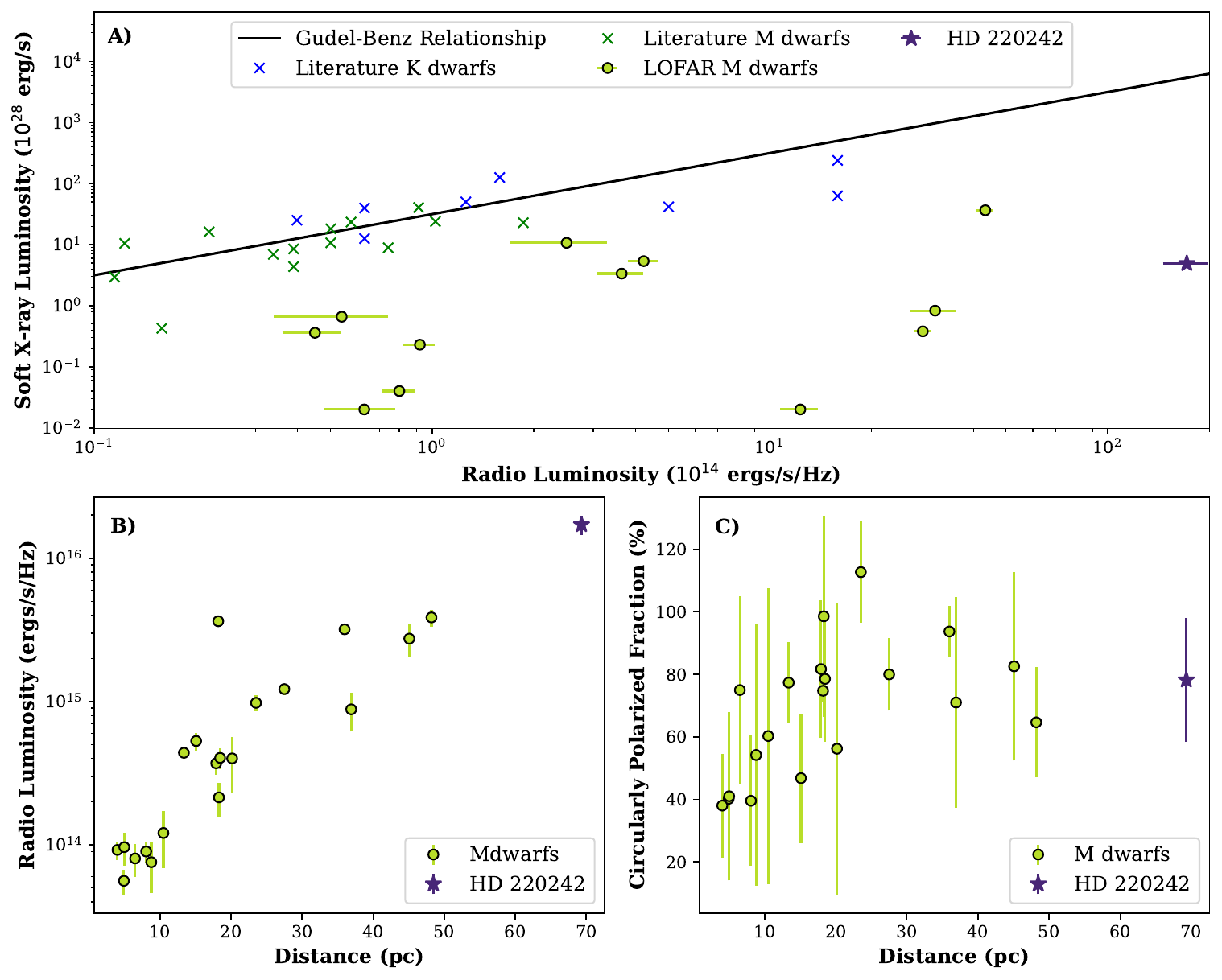}
    \caption{A)Radio luminosities of LOFAR M dwarf sample \citep{Callingham_population_2021} plotted against the ROSAT X-ray luminosities. The canonical G{\"u}del-Benz relationship \citep{benz_xray_1994} is plotted as a black line. Included for completeness are the M and K dwarfs originally used to calibrate the relationship \citep{gudel_radio_1992, Gudel_tight_1993}. While some members of the M-dwarf sample lie within the expected scatter on this relationship, the total X-ray luminosity from the HD~220242 system falls far below the expected value if it were following this relationship. B) Radio luminosities of the LOFAR M-dwarf sample compared to HD~220242. C) A comparison between the circularly polarized fraction of LOFAR M dwarfs and HD 220242.}
    \label{fig:binaryvsingle}
\end{figure*}

In addition to isolated main-sequence stars, the V-LoTSS sample contained multiple close M dwarf binaries. The mass for the secondary that is derived by our analysis could be split into a binary pair of two mid- or late-type M dwarfs rather than being wholly contained in a single body. Without any light from the secondary in the HPF spectra, we cannot conclusively state what type of star HD~220242B is; it could be a single chromospherically active early M dwarf, a single slowly rotating quiescent M dwarf with a planetary companion, or a binary pair of mid- or late-type M dwarfs. Failing to detect the light from the secondary is not totally surprising, as even a 4000\,K M dwarf would only be expected to have a flux ratio of 2.5\% in the HPF bandpass. However, as we hypothesize that the companion is the source of the radio emission, we compare characteristics like radio luminosity, circularly polarized fraction, and X-ray luminosity between the known M dwarf sample and HD~220422 in \autoref{fig:binaryvsingle} to search for trends in the data that may allow us to favor one possibility over the others.

As can be seen in \autoref{fig:binaryvsingle}, the fraction of circularly polarized emission varies widely across the M dwarf sample, and HD~220242 falls within the range. However, we can rule out the possibility of a closely interacting binary (i.e. RS CVn binaries) as these binaries follow the G{\"u}del-Benz relationship \citep{vedantham_peculiar_2022} and \autoref{fig:binaryvsingle} clearly shows that HD~220242 falls outside the observed scatter of the relationship. Of particular note is the luminosity of HD~220242: it is far more radio luminous than any member of the M dwarf sample. As we have eliminated the possibility of the emission coming from a background source, this could be caused by an extremely strong flare and may support a chromospherically active companion. The TESS observations of HD~220242 show no evidence of flaring, but the brightness of the primary means that a flare from the companion is unlikely to be detectable. Without being able to perform RV follow up directly on the companion, we cannot evaluate the possibility of a substellar companion to HD~220242B, though the low/solar metallicity does disfavor the possibility of a massive planetary companion \citep{Gan_metallicity_2025}.

%Regardless of the nature of the M dwarf companion, be it binary or single, or the emission mechanism, this particular system is significantly more luminous than any of its LOFAR M dwarf counterparts, suggesting a stronger radio engine than the ones in isolated M dwarfs.

We consider the case of a planet in a sub-Alfv\'enic orbit driving the radio emission here, following the formulation found in \citet{fitzmaurice_astrometry_2024}. A planet's orbit is sub-Alfv\'enic when the Alfv\'en Mach number (M$_A$) is less than unity \citep{Saur_magnetic_2013}:
\begin{equation}
    M_A=\frac{\Delta u}{u_A}<1,
\end{equation}
where $\Delta u$ is the velocity of the stellar wind in the rest frame of the planet and $u_A$ is the Alfv\'en speed. These quantities are dependent on the mass loss rate of the host star, as well as the average large-scale magnetic field strength. With no direct observations of HD~220242B, it is difficult to make any statements about these quantities; we draw reasonable values from the literature for similar early M~dwarfs. GJ~625 is an early M~dwarf (M2.5$\pm$0.4) with a known short period planet and detections of highly circularly polarized radio emission in multiple LOFAR pointings that are suggestive of sub-Alfv\'enic interactions \citep{Suarez_hades_2017,Koo_spectroscopic_2025}. Recent work considered mass loss rates between 0.68-25\,$\dot{M}_\odot$ and magnetic field strengths from 20-200\,G to assess the potential for these interactions to take place at the orbital distance of GJ~625b \citep{Koo_spectroscopic_2025} and we adopt those ranges for our calculation here. The power produced by sub-Alfv\'enic interactions is given by \citep{Saur_magnetic_2013, kavanagh_radio_2022}:
\begin{equation}
    P_{SA}=\pi^{1/2}R_{obs}^2B_w\rho_w^{1/2}\Delta u^2\sin^2\theta,
\end{equation}
where $R_{obs}$ is the effective radius of the perturbing body, $B_w$ is the magnetic field strength of the stellar wind at the orbital distance of the perturbing body, $\rho_w$ is the stellar wind density at that same point, and $\theta$ is the angle between the velocity and magnetic field vectors. We find that an object with an effective radius equal to the physical radius of Jupiter in a $\sim$1\,day period can reproduce the observed LOFAR emission at stellar mass loss rates of 8$\dot{M}_\odot$ or greater, even when accounting for the low efficiency of the Jupiter-Io interaction when converting to radio power \citep[$\sim$10$^{-3}$;][]{Turnpenney_exoplanet_2018, Saur_brown_2021}. With this brief calculation, we demonstrate that at least one of the mechanisms suggested to produce highly circularly polarized emission in isolated M~dwarfs is potentially capable of achieving the radio brightness observed from HD~220242.

Similar models have been used to investigate other M~dwarf systems with known short period planets, predicting that bursts of emission can reach $\sim1$\,mJy, with steady state flux densities around 10$\mu$Jy \citep{Turnpenney_exoplanet_2018}. To perform a similarly nuanced treatment and better understand properties of the radio emission other than power, we would require more information about HD~220242B than is currently available. This particular system is significantly more luminous than any of its LOFAR M dwarf counterparts, suggesting a stronger radio engine than the ones in isolated M dwarfs.

%Regardless of the nature of the M dwarf companion, be it binary or single, or the emission mechanism, this particular system is significantly more luminous than any of its LOFAR M dwarf counterparts, suggesting a stronger radio engine than the ones in isolated M dwarfs.

\subsubsection{Compact object companion}
%Highly polarized and periodic radio emission has previously been linked to neutron stars

We did not recover RVs from the companion in our spectroscopic binary analysis; this could indicate the companion is rapidly rotating or that it is a compact object that would not provide a second set of spectral lines. The derived mass of 0.619$\pm0.014$\:M$_\odot$ from our \texttt{orvara} analysis is consistent with that of a white dwarf ($\leq1.4$\:M$_\odot$), but excludes the possibility of the companion being a neutron star. Though highly polarized and periodic radio emission has primarily been linked to rotating neutron stars \citep{Bhattacharya_formatin_1991, Caleb_discovery_2022}, previous work has detected similar emission from magnetic white dwarfs \citep{Ayres_starcat_2010, Pelisoli_min_2023, deRuiter_white_2024}.

To evaluate the possibility of the detected companion being a WD, we first considered the evolution of a companion that formed at the same time as the primary. The MIST-based initial-final mass relation (IFMR) for WDs with initial stellar masses between 0.87 and 2.80\:M$_\odot$ is given as \citep{Cummings_IFMR_2018}:
\begin{equation}
    M_f = (0.080\pm0.0144) \times M_i + (0.489\pm0.030).
\end{equation}
For a WD mass of 0.62\:M$_\odot$, this gives a progenitor mass of 1.6$\pm0.5$\:M$_\odot$. The main sequence lifetime of a star of this mass is similar to that of the primary component of HD~220242, indicating a star of this mass is unlikely to have already evolved into a WD at the age of this system. If the companion were to have already undergone the asymptotic giant branch phase, it would have done so very recently and the resulting WD would still be extremely hot. The flux from the WD would present itself in UV and blue excess, but the GALEX photometry in \autoref{fig:sedfit} is consistent with that of a 6800\:K star. Additionally, HD~220242 was observed with STIS on the Hubble Space Telescope (HST) and the UV flux was consistent with model spectra of a 6800\:K star \citep{Ayres_starcat_2010}.

%It is possible for current companions to form separately at different times, then end up as binary pair through a 3-body interaction (look at steins thesis). However, such interactions are most likely to take place while stars are still contained within the clusters in which they formed. Cluster stars form within similar time frames therefore the resulting age difference is unlikely to be significant enough for the WD to have cooled sufficiently to no longer contribute any excess blue flux. %\textbf{it would take the sun several hubble times to interact with another star, find the calculation so that you can discard the idea of a significantly older WD companion.} 
While it is possible for binary pairs to destruct and reform during the evolution of young stellar clusters \citep{Parker_binaries_2009,Parker_binaries_2014}, observations of protostars reveal a decrease in binary fraction with age that has been attributed to dynamical interactions between stars \citep{Tobin_vla_2016}. However, binaries with separations $<50$\,au are considered ``hard'' and simulations have demonstrated they are unaffected by dynamical evolution in all but the most extreme cluster environments \citep{Parker_binaries_2009}. These simulations are supported by observations of binary fractions in young stellar associations, which are consistent with those found for field stars \citep{Elliot_search_2015}. The maximum separation between HD~220242 A \& B is 11.6\,au, indicating this binary would have been produced as a consequence of star formation processes and the companions should not differ significantly in age; not enough for a potential WD to have cooled sufficiently to no longer contribute excess blue flux.

Even if the companion were somehow a WD that is much older than the system, previous low-frequency radio detections have had a high degree of linear polarization. 
While recent work has found long period radio transients (LPRTs) to have high degrees of circular polarization \citep{Men_highly_2025}, magnetized WD are only one of several possible interpretations of this phenomena \citep{Rea_longperiod_2024}. We find this hypothesis unfavorable, but cannot completely discard the possibility.

\subsubsection{Stellar wind-magnetosphere interactions}
Auroral radio emission due to interactions between the energized particles of the solar wind and magnetic fields has been observed for all five magnetized solar system planets, and has been postulated as a possible way to detect exoplanets. This radio emission is driven by the ECMI mechanism, which would explain the high degree of circular polarization observed from the HD~220242 system. Using the Parker solar wind model, we approximate the 150\:MHz flux expected from the stellar wind of the primary interacting with the magnetic field of an M dwarf at the orbital distance of the secondary at the time of the LOFAR detection.

The LOFAR observation occurred on 2017 September 25, at which time the secondary was at an orbital distance of $\sim$10\:au based on our derived mass and orbital fit. In the Parker model, the wind density and velocity are set by the coronal temperature, surface gravity, and mass-loss rate of the host star. We derived the coronal temperature from the soft X-ray flux (F$_{\mathrm{x}}=4.11\times10^{5}\:\mathrm{erg\:s^{-1}\:cm^{-2}}$) at the surface of the star \citep{Johnstone_coronal_2015}: T$_c=0.11\mathrm{F_x}^{0.26}\times10^6=3.2 \times 10^6$\:K. The wind density and velocity at 10\,au are then $n(10\,{\rm au}) = (\dot{M}/\dot{M}_\odot)\times 0.034\,{\rm cm}^{-3}$ and $v(10\,{\rm au}) = 794\,{\rm km/s}$ respectively.

\begin{figure*}[htb]
    \centering
    \includegraphics[width=1.0\linewidth]{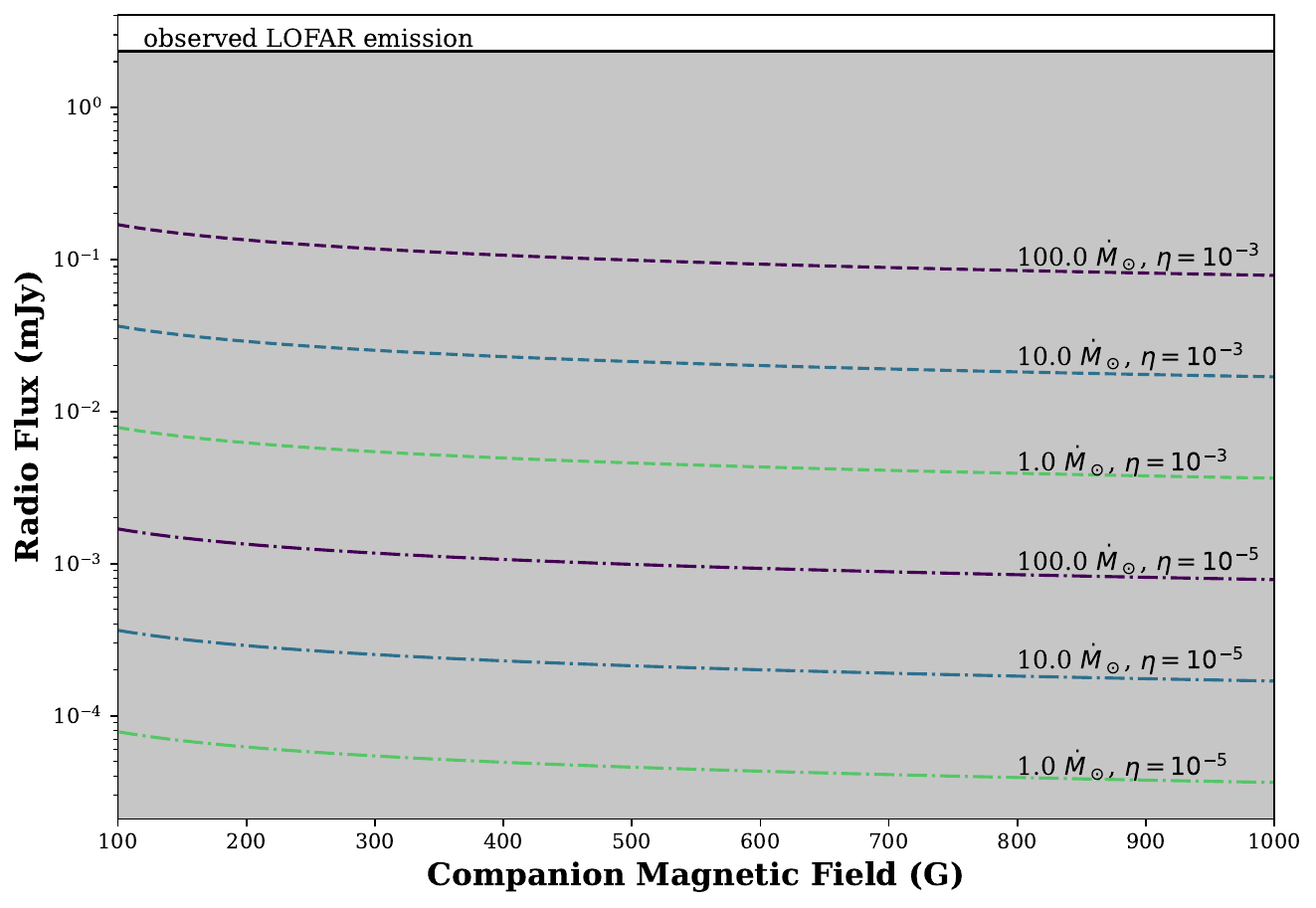}
    \caption{The expected radio flux due to an interaction between the stellar wind of the primary and the magnetosphere of the secondary as a function of the strength of the dipolar component of the secondary's magnetic field. Different line colors indicate different mass loss rates from the primary (1, 10, and 100 $\dot{M}_{\odot}$) and the different line styles indicate the conversion efficiency from ram pressure to radio flux (10$^{-5}$ and 10$^{-3}$). The grey shaded region indicates where the emission caused by such an interaction is not consistent with the LOFAR observation.}
    \label{fig:magneto}
\end{figure*}

From here, we assumed the dynamic pressure of the stellar wind dominates over the magnetic pressure it exerts --- true for the solar system, that the magnetosphere of the companion is dipolar, and that all of the intercepted energy flux from the stellar wind is converted into ECMI radiation from the companion (i.e. the intercepted \textbf{energy flux} is equal to the emitted radio luminosity and the efficiency $\eta=1$). This final assumption provides an absolute upper limit on the flux due to this interaction and we will revisit more likely assumptions later. For a dipolar magnetic field, the strength of the field goes as
\begin{equation}
    B(d) = B_0\left(\frac{R_c}{d_c}\right)^{3},
\end{equation}
where $B_0$ is the companion's surface magnetic field, $R_c$ is the radius of the companion, and $d_c$ is the distance from the companion. This then gives a magnetic pressure, $P_B$, of 
\begin{equation}
    P_{B}(d_c) = \frac{B_0^2}{8\pi}\left(\frac{R_c}{d_c}\right)^6.
\end{equation}
We equate the ram pressure, $P_R=0.5nm_pv^2$ where $m_p$ is the mass of a proton and $n$ is the stellar wind density, from the primary stellar wind with this magnetic pressure, solving for the distance from the companion at which the two are equal
\begin{equation}
    d_c = R_c\left(\frac{B_0^2}{8\pi P_R}\right)^{1/6}.
\end{equation}
This distance then becomes the radius for the area over which the ram pressure from the primary is intercepted by the secondary and converted into radio luminosity. This is a simple approximation for the area subtended by the magnetosphere that does not take into account the possibility of compression or more complicated magnetic field structures such as the one for Jupiter's magnetosphere, but provides a reasonable basis for a first-order estimate of the expected radio brightness. The final equation for expected radio flux is then
\begin{equation}
    F_\mathrm{Radio} = \eta\left(\frac{\pi d_c^2P_{R}v}{d^2\omega\Delta\nu}\right),
\end{equation}
where $d$ is the distance to the star, $\Delta\nu=2.8\times10^6B_0$\,Hz is the emission bandwidth, and $\omega$ is the beam solid angle of the radio emission, which we assume to be 1\,sr.
In \autoref{fig:magneto}, we explore what the expected radio flux in LOFAR would be for a range of different mass loss rates from the primary and different companion magnetic field strengths. The companion magnetic field strengths were chosen based on the results of \citet{Reiners_magnetism_2022}, who analyzed the magnetic fields of 292 M dwarfs; the majority of their sample had magnetic field strengths in the range of 200-1000\,G, though fields up to 8000\,G were also detected.

If we assume the conversion from ram pressure to radio flux is perfectly efficient, as we did above, a mass loss rate of 1\,$\dot{M}_{\odot}$ is sufficient to produce the emission observed by LOFAR (\autoref{fig:magneto}). However, studies of auroral emission in the Solar System have yielded empirical estimates of 10$^{-5}$ for the conversion efficiency from ram pressure to emitted radio power \citep{zarka_magnetically_2001}. At this efficiency, there is no reasonable mass loss rate that is able to replicate the observed emission. Even if we consider the higher conversion efficiency from Jupiter's interactions with Io ($\eta=$10$^{-3}$), there are no reasonable combinations of primary mass loss rate and secondary magnetic field strength that replicate the observed emission.

%We conducted this calculation using a series of assumptions and estimates that may or may not reflect the reality of the system. As such, we cannot confidently rule this mechanism out entirely as the driver of the observed emission, but suggest it is unlikely.

We conducted this calculation using a series of estimates that reasonably approximate the HD~220242 system, but caution that these calculations are approximations scaled from the Sun.

\section{Conclusion}

We present the dynamical mass of the companion to HD~220242 first reported in \citet{Hartmann_mass_2019}, having broken the sin\textit{i} degeneracy by including both intermediate Hipparcos astrometry and the Gaia-Hipparcos proper motion differences. Our spectroscopic and speckle observations along with archival STIS spectra lead us to strongly favor the hypothesis that the 0.619$\pm$0.014\:M$_\odot$ companion is a main-sequence M dwarf or a binary pair of two lower mass M dwarfs.

The source of the radio emission in the system most likely can be attributed to the companion. There are two possible ways to produce the emission: (1) interactions with the stellar wind of the primary or (2) direct emission, but the stellar wind-magnetosphere interaction is unlikely to be sufficient to drive the observed emission on its own, so we conclude some form of emission directly from the companion is necessary. We are able to demonstrate that a sub-Alfv\'enic interaction between HD~220242B and a Jupiter radius companion is potentially capable of powering the observed LOFAR emission, but caution that this is a non-exhaustive exploration of potential causes. We strongly favor emission from either a single or a binary pair of M dwarf companions over a WD, but additional observations sensitive to light from the secondary are required to distinguish between a single or binary pair. These additional observations could also be key to answering the question of why this particular system is so much more radio luminous than any of the other LOFAR M dwarf detections.

%% Also note that the akcnowlodgment environment does not support long amounts of text. If you have a lot of people and institutions to acknowledge, do not use this command. Instead, create a new \section{Acknowledgments}.
\newpage
\section{Acknowledgements}
We thank J. Villadsen for helpful suggestions in the preparation of the manuscript.

We acknowledge support from NASA XRP Grant 80NSSC24K0155.

%CIC
CIC acknowledges support by NASA Headquarters through an appointment to the NASA Postdoctoral Program at the Goddard Space Flight Center, administered by ORAU through a contract with NASA.

%JCG
JBM and JCG were supported by projects PID2023-147883NB-C22
funded by MCIN/AEI,CIPROM/2022/64, funded by the Generalitat Valenciana, and by the Astrophysics and High Energy Physics
programme by MCIN, with funding from European Union NextGenerationEU
(PRTR-C17.I1) and the Generalitat Valenciana through grant ASFAE/2022/018.

%MPT
MPT acknowledges financial support from the Severo Ochoa grant CEX2021-001131-S and from the Spanish grant PID2023-147883NB-C21, funded by MCIU/AEI/ 10.13039/501100011033, as well as support through ERDF/EU.

%HET
The Hobby-Eberly Telescope (HET) is a joint project of the University of Texas at Austin, the Pennsylvania State University, Ludwig-Maximilians-Universität München, and Georg-August-Universität Göttingen. The HET is named in honor of its principal benefactors, William P. Hobby and Robert E. Eberly.

%HPF
These results are based on observations obtained with the Habitable-zone Planet Finder Spectrograph on the HET. The HPF team was supported by NSF grants AST-1006676, AST-1126413, AST-1310885, AST-1517592, AST-1310875, AST-1910954, AST-1907622, AST-1909506, AST-2108493, AST-2108512, AST-2108569, AST-2108801, ATI-2009889, ATI-2009982, and the NASA Astrobiology Institute (NNA09DA76A) in the pursuit of precision radial velocities in the NIR. The HPF team was also supported by the Heising-Simons Foundation via grant 2017-0494.

%Gaia DR3
This work has made use of data from the European Space Agency (ESA) mission
{\it Gaia} (\url{https://www.cosmos.esa.int/gaia}), processed by the {\it Gaia}
Data Processing and Analysis Consortium (DPAC,
\url{https://www.cosmos.esa.int/web/gaia/dpac/consortium}). Funding for the DPAC
has been provided by national institutions, in particular the institutions
participating in the {\it Gaia} Multilateral Agreement.

%WIYN acknowledgment
WIYN is a joint facility of the University of Wisconsin-Madison, Indiana University, NSF's NOIRLab, the Pennsylvania State University, Purdue University, University of California-Irvine, and the University of Missouri. We thank the NEID Queue Observers and WIYN Observing Associates for their skillful execution of our NEID observations.

%Land acknowledgment
The authors are honored to be permitted to conduct astronomical research on Iolkam Du'ag (Kitt Peak), a mountain with particular significance to the Tohono O'odham. Data presented herein were obtained at the WIYN Observatory from telescope time allocated to NN-EXPLORE through the scientific partnership of NASA, the NSF, and NOIRLab.

% NESSI
Some of the observations in this paper made use of the NN-EXPLORE Exoplanet and Stellar Speckle Imager (NESSI). NESSI was funded by the NASA Exoplanet Exploration Program and the NASA Ames Research Center. NESSI was built at the Ames Research Center by Steve B. Howell, Nic Scott, Elliott P. Horch, and Emmett Quigley.

% ACI & Cyberlamp
Computations for this research were performed on the Pennsylvania State University’s Institute for Computational and Data Sciences Advanced CyberInfrastructure (ICDS-ACI).  This content is solely the responsibility of the authors and does not necessarily represent the views of the Institute for Computational and Data Sciences.

% CEHW 
The Center for Exoplanets and Habitable Worlds is supported by the Pennsylvania State University, the Eberly College of Science, and the Pennsylvania Space Grant Consortium. 

%ADS
This research has made use of the SIMBAD database, operated at CDS, Strasbourg, France, 
and NASA's Astrophysics Data System Bibliographic Services.

%MAST
Some/all of the data presented in this article were obtained from the Mikulski Archive for Space Telescopes (MAST) at the Space Telescope Science Institute. The specific observations analyzed can be accessed via \dataset[doi:10.17909/33tg-ps82]{https://doi.org/10.17909/33tg-ps82}.


\vspace{5mm}
\facilities{HPF/HET~10\,m, NESSI/WIYN~3.5\,m, \textit{Gaia}, HST/STIS, \textit{Hipparcos}}


\software{
\texttt{ARIADNE} \citep{Vines_ariadne_2022},
\texttt{astropy} \citep{robitaille_astropy_2013, astropy_collaboration_astropy_2018, astropy_2022},
\texttt{barycorrpy} \citep{kanodia_python_2018}, 
\texttt{dynesty} \citep{Speagle_dynesty_2020},
\texttt{emcee} \citep{foreman-mackey_emceee_2013},
\texttt{ipython} \citep{perez_ipython_2007},
\texttt{isochrones} \citep{isochrones},
\texttt{matplotlib} \citep{hunter_matplotlib_2007},
\texttt{numpy} \citep{harris_numpy_2020},
\texttt{orvara} \citep{brandt_orvara_2021},
\texttt{pandas} \citep{mckinney_data_2010, reback2020pandas},
\texttt{ptemcee} \citep{vousden_ptemcee_2016},
\texttt{RadVel} \citep{Fulton_radvel_2019},
\texttt{scipy} \citep{virtanen_scipy_2020},
\texttt{SERVAL} \citep{Zechmeister_spectrum_2018},
\texttt{TODCOR} \citep{Mazeh_todcor_1994}.
}




\bibliography{sample631,additional}{}

@ARTICLE{Suchkov_ROSAT_2003,
       author = {{Suchkov}, A.~A. and {Makarov}, V.~V. and {Voges}, W.},
        title = "{ROSAT View of Hipparcos F Stars}",
      journal = {\apj},
         year = 2003,
        month = oct,
       volume = {595},
       number = {2},
        pages = {1206-1221},
          doi = {10.1086/377472},
       adsurl = {https://ui.adsabs.harvard.edu/abs/2003ApJ...595.1206S}
}

@ARTICLE{Best_LOFAR_deepfields_2023,
       author = {{Best}, P.~N. and {Kondapally}, R. and {Williams}, W.~L. and {Cochrane}, R.~K. and {Duncan}, K.~J. and {Hale}, C.~L. and {Haskell}, P. and {Ma{\l}ek}, K. and {McCheyne}, I. and {Smith}, D.~J.~B. and {Wang}, L. and {Botteon}, A. and {Bonato}, M. and {Bondi}, M. and {Calistro Rivera}, G. and {Gao}, F. and {G{\"u}rkan}, G. and {Hardcastle}, M.~J. and {Jarvis}, M.~J. and {Mingo}, B. and {Miraghaei}, H. and {Morabito}, L.~K. and {Nisbet}, D. and {Prandoni}, I. and {R{\"o}ttgering}, H.~J.~A. and {Sabater}, J. and {Shimwell}, T. and {Tasse}, C. and {van Weeren}, R.},
        title = "{The LOFAR Two-metre Sky Survey: Deep Fields data release 1. V. Survey description, source classifications, and host galaxy properties}",
      journal = {\mnras},
         year = 2023,
        month = aug,
       volume = {523},
       number = {2},
        pages = {1729-1755},
          doi = {10.1093/mnras/stad1308},
archivePrefix = {arXiv},
       eprint = {2305.05782},
 primaryClass = {astro-ph.GA},
       adsurl = {https://ui.adsabs.harvard.edu/abs/2023MNRAS.523.1729B}
}

@ARTICLE{Petrov_widefield_2021,
       author = {{Petrov}, Leonid},
        title = "{The Wide-field VLBA Calibrator Survey: WFCS}",
      journal = {\aj},
         year = 2021,
        month = jan,
       volume = {161},
       number = {1},
          eid = {14},
        pages = {14},
          doi = {10.3847/1538-3881/abc4e1},
archivePrefix = {arXiv},
       eprint = {2008.09243},
 primaryClass = {astro-ph.IM},
       adsurl = {https://ui.adsabs.harvard.edu/abs/2021AJ....161...14P}
}

@ARTICLE{Cummings_IFMR_2018,
       author = {{Cummings}, Jeffrey D. and {Kalirai}, Jason S. and {Tremblay}, P. -E. and {Ramirez-Ruiz}, Enrico and {Choi}, Jieun},
        title = "{The White Dwarf Initial-Final Mass Relation for Progenitor Stars from 0.85 to 7.5 M $_{{\ensuremath{\odot}}}$}",
      journal = {\apj},
         year = 2018,
        month = oct,
       volume = {866},
       number = {1},
          eid = {21},
        pages = {21},
          doi = {10.3847/1538-4357/aadfd6},
archivePrefix = {arXiv},
       eprint = {1809.01673},
 primaryClass = {astro-ph.SR},
       adsurl = {https://ui.adsabs.harvard.edu/abs/2018ApJ...866...21C}
}

@ARTICLE{Johnstone_coronal_2015,
       author = {{Johnstone}, C.~P. and {G{\"u}del}, M.},
        title = "{The coronal temperatures of low-mass main-sequence stars}",
      journal = {\aap},
         year = 2015,
        month = jun,
       volume = {578},
          eid = {A129},
        pages = {A129},
          doi = {10.1051/0004-6361/201425283},
archivePrefix = {arXiv},
       eprint = {1505.00643},
 primaryClass = {astro-ph.SR},
       adsurl = {https://ui.adsabs.harvard.edu/abs/2015A&A...578A.129J}
}

@ARTICLE{Callingham_population_2021,
       author = {{Callingham}, J.~R. and {Vedantham}, H.~K. and {Shimwell}, T.~W. and {Pope}, B.~J.~S. and {Davis}, I.~E. and {Best}, P.~N. and {Hardcastle}, M.~J. and {R{\"o}ttgering}, H.~J.~A. and {Sabater}, J. and {Tasse}, C. and {van Weeren}, R.~J. and {Williams}, W.~L. and {Zarka}, P. and {de Gasperin}, F. and {Drabent}, A.},
        title = "{The population of M dwarfs observed at low radio frequencies}",
      journal = {Nature Astronomy},
         year = 2021,
        month = dec,
       volume = {5},
        pages = {1233-1239},
          doi = {10.1038/s41550-021-01483-0},
archivePrefix = {arXiv},
       eprint = {2110.03713},
 primaryClass = {astro-ph.SR},
       adsurl = {https://ui.adsabs.harvard.edu/abs/2021NatAs...5.1233C}
}

@ARTICLE{dulk_radio_1985,
       author = {{Dulk}, G.~A.},
        title = "{Radio emission from the sun and stars.}",
      journal = {\araa},
         year = 1985,
        month = jan,
       volume = {23},
        pages = {169-224},
          doi = {10.1146/annurev.aa.23.090185.001125},
       adsurl = {https://ui.adsabs.harvard.edu/abs/1985ARA&A..23..169D}
}

@ARTICLE{pope_tess_2021,
       author = {{Pope}, Benjamin J.~S. and {Callingham}, Joseph R. and {Feinstein}, Adina D. and {G{\"u}nther}, Maximilian N. and {Vedantham}, Harish K. and {Ansdell}, Megan and {Shimwell}, Timothy W.},
        title = "{The TESS View of LOFAR Radio-emitting Stars}",
      journal = {\apjl},
         year = 2021,
        month = sep,
       volume = {919},
       number = {1},
          eid = {L10},
        pages = {L10},
          doi = {10.3847/2041-8213/ac230c},
archivePrefix = {arXiv},
       eprint = {2110.04759},
 primaryClass = {astro-ph.SR},
       adsurl = {https://ui.adsabs.harvard.edu/abs/2021ApJ...919L..10P}
}

@ARTICLE{vedantham_peculiar_2022,
       author = {{Vedantham}, H.~K. and {Callingham}, J.~R. and {Shimwell}, T.~W. and {Benz}, A.~O. and {Hajduk}, M. and {Ray}, T.~P. and {Tasse}, C. and {Drabent}, A.},
        title = "{Peculiar Radio-X-Ray Relationship in Active Stars}",
      journal = {\apjl},
         year = 2022,
        month = feb,
       volume = {926},
       number = {2},
          eid = {L30},
        pages = {L30},
          doi = {10.3847/2041-8213/ac5115},
archivePrefix = {arXiv},
       eprint = {2201.12203},
 primaryClass = {astro-ph.SR},
       adsurl = {https://ui.adsabs.harvard.edu/abs/2022ApJ...926L..30V}
}

@INPROCEEDINGS{Mahadevan_hpf_2012,
       author = {{Mahadevan}, Suvrath and {Ramsey}, Lawrence and {Bender}, Chad and {Terrien}, Ryan and {Wright}, Jason T. and {Halverson}, Sam and {Hearty}, Fred and {Nelson}, Matt and {Burton}, Adam and {Redman}, Stephen and {Osterman}, Steven and {Diddams}, Scott and {Kasting}, James and {Endl}, Michael and {Deshpande}, Rohit},
        title = "{The habitable-zone planet finder: a stabilized fiber-fed NIR spectrograph for the Hobby-Eberly Telescope}",
    booktitle = {Ground-based and Airborne Instrumentation for Astronomy IV},
         year = 2012,
       editor = {{McLean}, Ian S. and {Ramsay}, Suzanne K. and {Takami}, Hideki},
       series = {Society of Photo-Optical Instrumentation Engineers (SPIE) Conference Series},
       volume = {8446},
        month = sep,
          eid = {84461S},
        pages = {84461S},
          doi = {10.1117/12.926102},
archivePrefix = {arXiv},
       eprint = {1209.1686},
 primaryClass = {astro-ph.EP},
       adsurl = {https://ui.adsabs.harvard.edu/abs/2012SPIE.8446E..1SM}
}

@INPROCEEDINGS{mahadevan_hpf_2014,
       author = {{Mahadevan}, Suvrath and {Ramsey}, Lawrence W. and {Terrien}, Ryan and {Halverson}, Samuel and {Roy}, Arpita and {Hearty}, Fred and {Levi}, Eric and {Stefansson}, Gudmundur K. and {Robertson}, Paul and {Bender}, Chad and {Schwab}, Chris and {Nelson}, Matt},
        title = "{The Habitable-zone Planet Finder: A status update on the development of a stabilized fiber-fed near-infrared spectrograph for the for the Hobby-Eberly telescope}",
    booktitle = {Ground-based and Airborne Instrumentation for Astronomy V},
         year = 2014,
       editor = {{Ramsay}, Suzanne K. and {McLean}, Ian S. and {Takami}, Hideki},
       series = {Society of Photo-Optical Instrumentation Engineers (SPIE) Conference Series},
       volume = {9147},
        month = jul,
          eid = {91471G},
        pages = {91471G},
          doi = {10.1117/12.2056417},
       adsurl = {https://ui.adsabs.harvard.edu/abs/2014SPIE.9147E..1GM}
}

@ARTICLE{Stefansson_temp_2016,
       author = {{Stefansson}, Gudmundur and {Hearty}, Frederick and {Robertson}, Paul and {Mahadevan}, Suvrath and {Anderson}, Tyler and {Levi}, Eric and {Bender}, Chad and {Nelson}, Matthew and {Monson}, Andrew and {Blank}, Basil and {Halverson}, Samuel and {Henderson}, Chuck and {Ramsey}, Lawrence and {Roy}, Arpita and {Schwab}, Christian and {Terrien}, Ryan},
        title = "{A Versatile Technique to Enable Sub-milli-Kelvin Instrument Stability for Precise Radial Velocity Measurements: Tests with the Habitable-zone Planet Finder}",
      journal = {\apj},
         year = 2016,
        month = dec,
       volume = {833},
       number = {2},
          eid = {175},
        pages = {175},
          doi = {10.3847/1538-4357/833/2/175},
archivePrefix = {arXiv},
       eprint = {1610.06216},
 primaryClass = {astro-ph.IM},
       adsurl = {https://ui.adsabs.harvard.edu/abs/2016ApJ...833..175S}
}

@ARTICLE{Shetrone_queue_2007,
       author = {{Shetrone}, Matthew and {Cornell}, Mark E. and {Fowler}, James R. and {Gaffney}, Niall and {Laws}, Benjamin and {Mader}, Jeff and {Mason}, Cloud and {Odewahn}, Stephen and {Roman}, Brian and {Rostopchin}, Sergey and {Schneider}, Donald P. and {Umbarger}, James and {Westfall}, Amy},
        title = "{Ten Year Review of Queue Scheduling of the Hobby-Eberly Telescope}",
      journal = {\pasp},
         year = 2007,
        month = may,
       volume = {119},
       number = {855},
        pages = {556-566},
          doi = {10.1086/519291},
archivePrefix = {arXiv},
       eprint = {0705.3889},
 primaryClass = {astro-ph},
       adsurl = {https://ui.adsabs.harvard.edu/abs/2007PASP..119..556S}
}

@ARTICLE{stefansson_neptune_2023,
       author = {{Stef{\'a}nsson}, Gu{\dh}mundur and {Mahadevan}, Suvrath and {Miguel}, Yamila and {Robertson}, Paul and {Delamer}, Megan and {Kanodia}, Shubham and {Ca{\~n}as}, Caleb I. and {Winn}, Joshua N. and {Ninan}, Joe P. and {Terrien}, Ryan C. and {Holcomb}, Rae and {Ford}, Eric B. and {Zawadzki}, Brianna and {Bowler}, Brendan P. and {Bender}, Chad F. and {Cochran}, William D. and {Diddams}, Scott and {Endl}, Michael and {Fredrick}, Connor and {Halverson}, Samuel and {Hearty}, Fred and {Hill}, Gary J. and {Lin}, Andrea S.~J. and {Metcalf}, Andrew J. and {Monson}, Andrew and {Ramsey}, Lawrence and {Roy}, Arpita and {Schwab}, Christian and {Wright}, Jason T. and {Zeimann}, Gregory},
        title = "{A Neptune-mass exoplanet in close orbit around a very low-mass star challenges formation models}",
      journal = {Science},
         year = 2023,
        month = dec,
       volume = {382},
       number = {6674},
        pages = {1031-1035},
          doi = {10.1126/science.abo0233},
archivePrefix = {arXiv},
       eprint = {2303.13321},
 primaryClass = {astro-ph.EP},
       adsurl = {https://ui.adsabs.harvard.edu/abs/2023Sci...382.1031S}
}

@ARTICLE{wielebinski_history_2012,
       author = {{Wielebinski}, Richard},
        title = "{A history of radio astronomy polarisation measurements}",
      journal = {Journal of Astronomical History and Heritage},
         year = 2012,
        month = jul,
       volume = {15},
       number = {2},
        pages = {76-95},
       adsurl = {https://ui.adsabs.harvard.edu/abs/2012JAHH...15...76W}
}

@ARTICLE{Vines_ariadne_2022,
       author = {{Vines}, Jose I. and {Jenkins}, James S.},
        title = "{ARIADNE: measuring accurate and precise stellar parameters through SED fitting}",
      journal = {\mnras},
         year = 2022,
        month = jun,
       volume = {513},
       number = {2},
        pages = {2719-2731},
          doi = {10.1093/mnras/stac956},
archivePrefix = {arXiv},
       eprint = {2204.03769},
 primaryClass = {astro-ph.SR},
       adsurl = {https://ui.adsabs.harvard.edu/abs/2022MNRAS.513.2719V}
}

@dataset{kurucz_1993,
       author = {{Kurucz}, R.~L.},
        title = "{VizieR Online Data Catalog: Model Atmospheres (Kurucz, 1979)}",
 howpublished = {VizieR On-line Data Catalog: VI/39.  Originally published in: 1979ApJS...40....1K},
         year = 1993,
        month = oct,
          eid = {VI/39},
       adsurl = {https://ui.adsabs.harvard.edu/abs/1993yCat.6039....0K}
}

@INPROCEEDINGS{Castelli_2003,
       author = {{Castelli}, F. and {Kurucz}, R.~L.},
        title = "{New Grids of ATLAS9 Model Atmospheres}",
    booktitle = {Modelling of Stellar Atmospheres},
         year = 2003,
       editor = {{Piskunov}, N. and {Weiss}, W.~W. and {Gray}, D.~F.},
       series = {IAU Symposium},
       volume = {210},
        month = jan,
        pages = {A20},
          doi = {10.48550/arXiv.astro-ph/0405087},
archivePrefix = {arXiv},
       eprint = {astro-ph/0405087},
 primaryClass = {astro-ph},
       adsurl = {https://ui.adsabs.harvard.edu/abs/2003IAUS..210P.A20C}
}

@ARTICLE{brandt_orvara_2021,
       author = {{Brandt}, Timothy D. and {Dupuy}, Trent J. and {Li}, Yiting and {Brandt}, G. Mirek and {Zeng}, Yunlin and {Michalik}, Daniel and {Bardalez Gagliuffi}, Daniella C. and {Raposo-Pulido}, Virginia},
        title = "{orvara: An Efficient Code to Fit Orbits Using Radial Velocity, Absolute, and/or Relative Astrometry}",
      journal = {\aj},
         year = 2021,
        month = nov,
       volume = {162},
       number = {5},
          eid = {186},
        pages = {186},
          doi = {10.3847/1538-3881/ac042e},
archivePrefix = {arXiv},
       eprint = {2105.11671},
 primaryClass = {astro-ph.IM},
       adsurl = {https://ui.adsabs.harvard.edu/abs/2021AJ....162..186B}
}

@ARTICLE{brandt_HGCA_2018,
       author = {{Brandt}, Timothy D.},
        title = "{The Hipparcos-Gaia Catalog of Accelerations}",
      journal = {\apjs},
         year = 2018,
        month = dec,
       volume = {239},
       number = {2},
          eid = {31},
        pages = {31},
          doi = {10.3847/1538-4365/aaec06},
archivePrefix = {arXiv},
       eprint = {1811.07283},
 primaryClass = {astro-ph.SR},
       adsurl = {https://ui.adsabs.harvard.edu/abs/2018ApJS..239...31B}
}

@ARTICLE{vousden_ptemcee_2016,
       author = {{Vousden}, W.~D. and {Farr}, W.~M. and {Mandel}, I.},
        title = "{Dynamic temperature selection for parallel tempering in Markov chain Monte Carlo simulations}",
      journal = {\mnras},
         year = 2016,
        month = jan,
       volume = {455},
       number = {2},
        pages = {1919-1937},
          doi = {10.1093/mnras/stv2422},
archivePrefix = {arXiv},
       eprint = {1501.05823},
 primaryClass = {astro-ph.IM},
       adsurl = {https://ui.adsabs.harvard.edu/abs/2016MNRAS.455.1919V}
}

@ARTICLE{foreman-mackey_emceee_2013,
       author = {{Foreman-Mackey}, Daniel and {Hogg}, David W. and {Lang}, Dustin and {Goodman}, Jonathan},
        title = "{emcee: The MCMC Hammer}",
      journal = {\pasp},
         year = 2013,
        month = mar,
       volume = {125},
       number = {925},
        pages = {306},
          doi = {10.1086/670067},
archivePrefix = {arXiv},
       eprint = {1202.3665},
 primaryClass = {astro-ph.IM},
       adsurl = {https://ui.adsabs.harvard.edu/abs/2013PASP..125..306F}
}

@ARTICLE{Shimwell_lofar_2017,
       author = {{Shimwell}, T.~W. and {R{\"o}ttgering}, H.~J.~A. and {Best}, P.~N. and {Williams}, W.~L. and {Dijkema}, T.~J. and {de Gasperin}, F. and {Hardcastle}, M.~J. and {Heald}, G.~H. and {Hoang}, D.~N. and {Horneffer}, A. and {Intema}, H. and {Mahony}, E.~K. and {Mandal}, S. and {Mechev}, A.~P. and {Morabito}, L. and {Oonk}, J.~B.~R. and {Rafferty}, D. and {Retana-Montenegro}, E. and {Sabater}, J. and {Tasse}, C. and {van Weeren}, R.~J. and {Br{\"u}ggen}, M. and {Brunetti}, G. and {Chy{\.z}y}, K.~T. and {Conway}, J.~E. and {Haverkorn}, M. and {Jackson}, N. and {Jarvis}, M.~J. and {McKean}, J.~P. and {Miley}, G.~K. and {Morganti}, R. and {White}, G.~J. and {Wise}, M.~W. and {van Bemmel}, I.~M. and {Beck}, R. and {Brienza}, M. and {Bonafede}, A. and {Calistro Rivera}, G. and {Cassano}, R. and {Clarke}, A.~O. and {Cseh}, D. and {Deller}, A. and {Drabent}, A. and {van Driel}, W. and {Engels}, D. and {Falcke}, H. and {Ferrari}, C. and {Fr{\"o}hlich}, S. and {Garrett}, M.~A. and {Harwood}, J.~J. and {Heesen}, V. and {Hoeft}, M. and {Horellou}, C. and {Israel}, F.~P. and {Kapi{\'n}ska}, A.~D. and {Kunert-Bajraszewska}, M. and {McKay}, D.~J. and {Mohan}, N.~R. and {Orr{\'u}}, E. and {Pizzo}, R.~F. and {Prandoni}, I. and {Schwarz}, D.~J. and {Shulevski}, A. and {Sipior}, M. and {Smith}, D.~J.~B. and {Sridhar}, S.~S. and {Steinmetz}, M. and {Stroe}, A. and {Varenius}, E. and {van der Werf}, P.~P. and {Zensus}, J.~A. and {Zwart}, J.~T.~L.},
        title = "{The LOFAR Two-metre Sky Survey. I. Survey description and preliminary data release}",
      journal = {\aap},
         year = 2017,
        month = feb,
       volume = {598},
          eid = {A104},
        pages = {A104},
          doi = {10.1051/0004-6361/201629313},
archivePrefix = {arXiv},
       eprint = {1611.02700},
 primaryClass = {astro-ph.IM},
       adsurl = {https://ui.adsabs.harvard.edu/abs/2017A&A...598A.104S}
}

@ARTICLE{Toet_coherent_2021,
       author = {{Toet}, S.~E.~B. and {Vedantham}, H.~K. and {Callingham}, J.~R. and {Veken}, K.~C. and {Shimwell}, T.~W. and {Zarka}, P. and {R{\"o}ttgering}, H.~J.~A. and {Drabent}, A.},
        title = "{Coherent radio emission from a population of RS Canum Venaticorum systems}",
      journal = {\aap},
         year = 2021,
        month = oct,
       volume = {654},
          eid = {A21},
        pages = {A21},
          doi = {10.1051/0004-6361/202141163},
archivePrefix = {arXiv},
       eprint = {2107.06690},
 primaryClass = {astro-ph.SR},
       adsurl = {https://ui.adsabs.harvard.edu/abs/2021A&A...654A..21T}
}

@ARTICLE{Ayres_starcat_2010,
       author = {{Ayres}, Thomas R.},
        title = "{StarCAT: A Catalog of Space Telescope Imaging Spectrograph Ultraviolet Echelle Spectra of Stars}",
      journal = {\apjs},
         year = 2010,
        month = mar,
       volume = {187},
       number = {1},
        pages = {149-171},
          doi = {10.1088/0067-0049/187/1/149},
       adsurl = {https://ui.adsabs.harvard.edu/abs/2010ApJS..187..149A}
}

@ARTICLE{Pelisoli_min_2023,
       author = {{Pelisoli}, Ingrid and {Marsh}, T.~R. and {Buckley}, David A.~H. and {Heywood}, I. and {Potter}, Stephen. B. and {Schwope}, Axel and {Brink}, Jaco and {Standke}, Annie and {Woudt}, P.~A. and {Parsons}, S.~G. and {Green}, M.~J. and {Kepler}, S.~O. and {Munday}, James and {Romero}, A.~D. and {Breedt}, E. and {Brown}, A.~J. and {Dhillon}, V.~S. and {Dyer}, M.~J. and {Kerry}, P. and {Littlefair}, S.~P. and {Sahman}, D.~I. and {Wild}, J.~F.},
        title = "{A 5.3-min-period pulsing white dwarf in a binary detected from radio to X-rays}",
      journal = {Nature Astronomy},
         year = 2023,
        month = aug,
       volume = {7},
        pages = {931-942},
          doi = {10.1038/s41550-023-01995-x},
archivePrefix = {arXiv},
       eprint = {2306.09272},
 primaryClass = {astro-ph.SR},
       adsurl = {https://ui.adsabs.harvard.edu/abs/2023NatAs...7..931P}
}

@ARTICLE{Bhattacharya_formatin_1991,
       author = {{Bhattacharya}, D. and {van den Heuvel}, E.~P.~J.},
        title = "{Formation and evolution of binary and millisecond radio pulsars}",
      journal = {\physrep},
         year = 1991,
        month = jan,
       volume = {203},
       number = {1-2},
        pages = {1-124},
          doi = {10.1016/0370-1573(91)90064-S},
       adsurl = {https://ui.adsabs.harvard.edu/abs/1991PhR...203....1B}
}

@ARTICLE{Caleb_discovery_2022,
       author = {{Caleb}, Manisha and {Heywood}, Ian and {Rajwade}, Kaustubh and {Malenta}, Mateusz and {Stappers}, Benjamin Willem and {Barr}, Ewan and {Chen}, Weiwei and {Morello}, Vincent and {Sanidas}, Sotiris and {van den Eijnden}, Jakob and {Kramer}, Michael and {Buckley}, David and {Brink}, Jaco and {Motta}, Sara Elisa and {Woudt}, Patrick and {Weltevrede}, Patrick and {Jankowski}, Fabian and {Surnis}, Mayuresh and {Buchner}, Sarah and {Bezuidenhout}, Mechiel Christiaan and {Driessen}, Laura Nicole and {Fender}, Rob},
        title = "{Discovery of a radio-emitting neutron star with an ultra-long spin period of 76 s}",
      journal = {Nature Astronomy},
         year = 2022,
        month = may,
       volume = {6},
        pages = {828-836},
          doi = {10.1038/s41550-022-01688-x},
archivePrefix = {arXiv},
       eprint = {2206.01346},
 primaryClass = {astro-ph.HE},
       adsurl = {https://ui.adsabs.harvard.edu/abs/2022NatAs...6..828C}
}

@ARTICLE{Weiler_catalog_1983,
       author = {{Weiler}, K.~W. and {de Pater}, I.},
        title = "{A catalog of high accuracy polarization measurements.}",
      journal = {\apjs},
         year = 1983,
        month = jul,
       volume = {52},
        pages = {293-327},
          doi = {10.1086/190869},
       adsurl = {https://ui.adsabs.harvard.edu/abs/1983ApJS...52..293W}
}

@ARTICLE{Rayner_radiocircular_2000,
       author = {{Rayner}, D.~P. and {Norris}, R.~P. and {Sault}, R.~J.},
        title = "{Radio circular polarization of active galaxies}",
      journal = {\mnras},
         year = 2000,
        month = dec,
       volume = {319},
       number = {2},
        pages = {484-496},
          doi = {10.1046/j.1365-8711.2000.03854.x},
       adsurl = {https://ui.adsabs.harvard.edu/abs/2000MNRAS.319..484R}
}

@ARTICLE{Agudo_polarized_2022,
       author = {{Agudo}, Iv{\'a}n and {Thum}, Clemens},
        title = "{The Polarized Emission of AGN at Millimeter Wavelengths as Seen by POLAMI}",
      journal = {Galaxies},
         year = 2022,
        month = aug,
       volume = {10},
       number = {4},
          eid = {87},
        pages = {87},
          doi = {10.3390/galaxies10040087},
       adsurl = {https://ui.adsabs.harvard.edu/abs/2022Galax..10...87A}
}

@ARTICLE{Farrell_possibility_1999,
       author = {{Farrell}, W.~M. and {Desch}, M.~D. and {Zarka}, P.},
        title = "{On the possibility of coherent cyclotron emission from extrasolar planets}",
      journal = {\jgr},
         year = 1999,
        month = jun,
       volume = {104},
       number = {E6},
        pages = {14025-14032},
          doi = {10.1029/1998JE900050},
       adsurl = {https://ui.adsabs.harvard.edu/abs/1999JGR...10414025F}
}

@ARTICLE{Parker_binaries_2014,
       author = {{Parker}, Richard J. and {Meyer}, Michael R.},
        title = "{Binaries in the field: fossils of the star formation process?}",
      journal = {\mnras},
         year = 2014,
        month = aug,
       volume = {442},
       number = {4},
        pages = {3722-3736},
          doi = {10.1093/mnras/stu1101},
archivePrefix = {arXiv},
       eprint = {1406.0844},
 primaryClass = {astro-ph.SR},
       adsurl = {https://ui.adsabs.harvard.edu/abs/2014MNRAS.442.3722P}
}

@ARTICLE{Tobin_vla_2016,
       author = {{Tobin}, John J. and {Looney}, Leslie W. and {Li}, Zhi-Yun and {Chandler}, Claire J. and {Dunham}, Michael M. and {Segura-Cox}, Dominique and {Sadavoy}, Sarah I. and {Melis}, Carl and {Harris}, Robert J. and {Kratter}, Kaitlin and {Perez}, Laura},
        title = "{The VLA Nascent Disk and Multiplicity Survey of Perseus Protostars (VANDAM). II. Multiplicity of Protostars in the Perseus Molecular Cloud}",
      journal = {\apj},
         year = 2016,
        month = feb,
       volume = {818},
       number = {1},
          eid = {73},
        pages = {73},
          doi = {10.3847/0004-637X/818/1/73},
archivePrefix = {arXiv},
       eprint = {1601.00692},
 primaryClass = {astro-ph.SR},
       adsurl = {https://ui.adsabs.harvard.edu/abs/2016ApJ...818...73T}
}

@ARTICLE{Parker_binaries_2009,
       author = {{Parker}, Richard J. and {Goodwin}, Simon P. and {Kroupa}, Pavel and {Kouwenhoven}, M.~B.~N.},
        title = "{Do binaries in clusters form in the same way as in the field?}",
      journal = {\mnras},
         year = 2009,
        month = aug,
       volume = {397},
       number = {3},
        pages = {1577-1586},
          doi = {10.1111/j.1365-2966.2009.15032.x},
archivePrefix = {arXiv},
       eprint = {0905.2140},
 primaryClass = {astro-ph.GA},
       adsurl = {https://ui.adsabs.harvard.edu/abs/2009MNRAS.397.1577P}
}

@ARTICLE{Elliot_search_2015,
       author = {{Elliott}, P. and {Hu{\'e}lamo}, N. and {Bouy}, H. and {Bayo}, A. and {Melo}, C.~H.~F. and {Torres}, C.~A.~O. and {Sterzik}, M.~F. and {Quast}, G.~R. and {Chauvin}, G. and {Barrado}, D.},
        title = "{Search for associations containing young stars (SACY). VI. Is multiplicity universal? Stellar multiplicity in the range 3-1000 au from adaptive-optics observations}",
      journal = {\aap},
         year = 2015,
        month = aug,
       volume = {580},
          eid = {A88},
        pages = {A88},
          doi = {10.1051/0004-6361/201525794},
archivePrefix = {arXiv},
       eprint = {1505.07837},
 primaryClass = {astro-ph.SR},
       adsurl = {https://ui.adsabs.harvard.edu/abs/2015A&A...580A..88E}
}

@ARTICLE{bailer-jones_estimating_2021,
       author = {{Bailer-Jones}, C.~A.~L. and {Rybizki}, J. and {Fouesneau}, M. and {Demleitner}, M. and {Andrae}, R.},
        title = "{Estimating Distances from Parallaxes. V. Geometric and Photogeometric Distances to 1.47 Billion Stars in Gaia Early Data Release 3}",
      journal = {\aj},
         year = 2021,
        month = mar,
       volume = {161},
       number = {3},
          eid = {147},
        pages = {147},
          doi = {10.3847/1538-3881/abd806},
archivePrefix = {arXiv},
       eprint = {2012.05220},
 primaryClass = {astro-ph.SR},
       adsurl = {https://ui.adsabs.harvard.edu/abs/2021AJ....161..147B}
}

@ARTICLE{wright_wide-field_2010,
       author = {{Wright}, Edward L. and {Eisenhardt}, Peter R.~M. and {Mainzer}, Amy K. and {Ressler}, Michael E. and {Cutri}, Roc M. and {Jarrett}, Thomas and {Kirkpatrick}, J. Davy and {Padgett}, Deborah and {McMillan}, Robert S. and {Skrutskie}, Michael and {Stanford}, S.~A. and {Cohen}, Martin and {Walker}, Russell G. and {Mather}, John C. and {Leisawitz}, David and {Gautier}, Thomas N., III and {McLean}, Ian and {Benford}, Dominic and {Lonsdale}, Carol J. and {Blain}, Andrew and {Mendez}, Bryan and {Irace}, William R. and {Duval}, Valerie and {Liu}, Fengchuan and {Royer}, Don and {Heinrichsen}, Ingolf and {Howard}, Joan and {Shannon}, Mark and {Kendall}, Martha and {Walsh}, Amy L. and {Larsen}, Mark and {Cardon}, Joel G. and {Schick}, Scott and {Schwalm}, Mark and {Abid}, Mohamed and {Fabinsky}, Beth and {Naes}, Larry and {Tsai}, Chao-Wei},
        title = "{The Wide-field Infrared Survey Explorer (WISE): Mission Description and Initial On-orbit Performance}",
      journal = {\aj},
         year = 2010,
        month = dec,
       volume = {140},
       number = {6},
        pages = {1868-1881},
          doi = {10.1088/0004-6256/140/6/1868},
archivePrefix = {arXiv},
       eprint = {1008.0031},
 primaryClass = {astro-ph.IM},
       adsurl = {https://ui.adsabs.harvard.edu/abs/2010AJ....140.1868W}
}

@ARTICLE{stassun_tess_2018,
       author = {{Stassun}, Keivan G. and {Oelkers}, Ryan J. and {Pepper}, Joshua and {Paegert}, Martin and {De Lee}, Nathan and {Torres}, Guillermo and {Latham}, David W. and {Charpinet}, St{\'e}phane and {Dressing}, Courtney D. and {Huber}, Daniel and {Kane}, Stephen R. and {L{\'e}pine}, S{\'e}bastien and {Mann}, Andrew and {Muirhead}, Philip S. and {Rojas-Ayala}, B{\'a}rbara and {Silvotti}, Roberto and {Fleming}, Scott W. and {Levine}, Al and {Plavchan}, Peter},
        title = "{The TESS Input Catalog and Candidate Target List}",
      journal = {\aj},
         year = 2018,
        month = sep,
       volume = {156},
       number = {3},
          eid = {102},
        pages = {102},
          doi = {10.3847/1538-3881/aad050},
archivePrefix = {arXiv},
       eprint = {1706.00495},
 primaryClass = {astro-ph.EP},
       adsurl = {https://ui.adsabs.harvard.edu/abs/2018AJ....156..102S}
}

@BOOK{cutri_2mass_2003,
       author = {{Cutri}, R.~M. and {Skrutskie}, M.~F. and {van Dyk}, S. and {Beichman}, C.~A. and {Carpenter}, J.~M. and {Chester}, T. and {Cambresy}, L. and {Evans}, T. and {Fowler}, J. and {Gizis}, J. and {Howard}, E. and {Huchra}, J. and {Jarrett}, T. and {Kopan}, E.~L. and {Kirkpatrick}, J.~D. and {Light}, R.~M. and {Marsh}, K.~A. and {McCallon}, H. and {Schneider}, S. and {Stiening}, R. and {Sykes}, M. and {Weinberg}, M. and {Wheaton}, W.~A. and {Wheelock}, S. and {Zacarias}, N.},
        title = "{2MASS All Sky Catalog of point sources.}",
         year = 2003,
       adsurl = {https://ui.adsabs.harvard.edu/abs/2003tmc..book.....C}
}

@ARTICLE{gaia_collaboration_gaia_2022,
       author = {{Gaia Collaboration} and {Klioner}, S.~A. and {Lindegren}, L. and {Mignard}, F. and {Hern{\'a}ndez}, J. and {Ramos-Lerate}, M. and {Bastian}, U. and {Biermann}, M. and {Bombrun}, A. and {de Torres}, A. and {Gerlach}, E. and {Geyer}, R. and {Hilger}, T. and {Hobbs}, D. and {Lammers}, U.~L. and {McMillan}, P.~J. and {Steidelm{\"u}ller}, H. and {Teyssier}, D. and {Raiteri}, C.~M. and {Bartolom{\'e}}, S. and {Bernet}, M. and {Casta{\~n}eda}, J. and {Clotet}, M. and {Davidson}, M. and {Fabricius}, C. and {Garralda Torres}, N. and {Gonz{\'a}lez-Vidal}, J.~J. and {Portell}, J. and {Rowell}, N. and {Torra}, F. and {Torra}, J. and {Brown}, A.~G.~A. and {Vallenari}, A. and {Prusti}, T. and {de Bruijne}, J.~H.~J. and {Arenou}, F. and {Babusiaux}, C. and {Creevey}, O.~L. and {Ducourant}, C. and {Evans}, D.~W. and {Eyer}, L. and {Guerra}, R. and {Hutton}, A. and {Jordi}, C. and {Luri}, X. and {Panem}, C. and {Pourbaix}, D. and {Randich}, S. and {Sartoretti}, P. and {Soubiran}, C. and {Tanga}, P. and {Walton}, N.~A. and {Bailer-Jones}, C.~A.~L. and {Drimmel}, R. and {Jansen}, F. and {Katz}, D. and {Lattanzi}, M.~G. and {van Leeuwen}, F. and {Bakker}, J. and {Cacciari}, C. and {De Angeli}, F. and {Fouesneau}, M. and {Fr{\'e}mat}, Y. and {Galluccio}, L. and {Guerrier}, A. and {Heiter}, U. and {Masana}, E. and {Messineo}, R. and {Mowlavi}, N. and {Nicolas}, C. and {Nienartowicz}, K. and {Pailler}, F. and {Panuzzo}, P. and {Riclet}, F. and {Roux}, W. and {Seabroke}, G.~M. and {Sordo}, R. and {Th{\'e}venin}, F. and {Gracia-Abril}, G. and {Altmann}, M. and {Andrae}, R. and {Audard}, M. and {Bellas-Velidis}, I. and {Benson}, K. and {Berthier}, J. and {Blomme}, R. and {Burgess}, P.~W. and {Busonero}, D. and {Busso}, G. and {C{\'a}novas}, H. and {Carry}, B. and {Cellino}, A. and {Cheek}, N. and {Clementini}, G. and {Damerdji}, Y. and {de Teodoro}, P. and {Nu{\~n}ez Campos}, M. and {Delchambre}, L. and {Dell'Oro}, A. and {Esquej}, P. and {Fern{\'a}ndez-Hern{\'a}ndez}, J. and {Fraile}, E. and {Garabato}, D. and {Garc{\'\i}a-Lario}, P. and {Gosset}, E. and {Haigron}, R. and {Halbwachs}, J. -L. and {Hambly}, N.~C. and {Harrison}, D.~L. and {Hestroffer}, D. and {Hodgkin}, S.~T. and {Holl}, B. and {Jan{\ss}en}, K. and {Jevardat de Fombelle}, G. and {Jordan}, S. and {Krone-Martins}, A. and {Lanzafame}, A.~C. and {L{\"o}ffler}, W. and {Marchal}, O. and {Marrese}, P.~M. and {Moitinho}, A. and {Muinonen}, K. and {Osborne}, P. and {Pancino}, E. and {Pauwels}, T. and {Recio-Blanco}, A. and {Reyl{\'e}}, C. and {Riello}, M. and {Rimoldini}, L. and {Roegiers}, T. and {Rybizki}, J. and {Sarro}, L.~M. and {Siopis}, C. and {Smith}, M. and {Sozzetti}, A. and {Utrilla}, E. and {van Leeuwen}, M. and {Abbas}, U. and {{\'A}brah{\'a}m}, P. and {Abreu Aramburu}, A. and {Aerts}, C. and {Aguado}, J.~J. and {Ajaj}, M. and {Aldea-Montero}, F. and {Altavilla}, G. and {{\'A}lvarez}, M.~A. and {Alves}, J. and {Anderson}, R.~I. and {Anglada Varela}, E. and {Antoja}, T. and {Baines}, D. and {Baker}, S.~G. and {Balaguer-N{\'u}{\~n}ez}, L. and {Balbinot}, E. and {Balog}, Z. and {Barache}, C. and {Barbato}, D. and {Barros}, M. and {Barstow}, M.~A. and {Bassilana}, J. -L. and {Bauchet}, N. and {Becciani}, U. and {Bellazzini}, M. and {Berihuete}, A. and {Bertone}, S. and {Bianchi}, L. and {Binnenfeld}, A. and {Blanco-Cuaresma}, S. and {Boch}, T. and {Bossini}, D. and {Bouquillon}, S. and {Bragaglia}, A. and {Bramante}, L. and {Breedt}, E. and {Bressan}, A. and {Brouillet}, N. and {Brugaletta}, E. and {Bucciarelli}, B. and {Burlacu}, A. and {Butkevich}, A.~G. and {Buzzi}, R. and {Caffau}, E. and {Cancelliere}, R. and {Cantat-Gaudin}, T. and {Carballo}, R. and {Carlucci}, T. and {Carnerero}, M.~I. and {Carrasco}, J.~M. and {Casamiquela}, L. and {Castellani}, M. and {Castro-Ginard}, A. and {Chaoul}, L. and {Charlot}, P. and {Chemin}, L. and {Chiaramida}, V. and {Chiavassa}, A. and {Chornay}, N. and {Comoretto}, G. and {Contursi}, G. and {Cooper}, W.~J. and {Cornez}, T. and {Cowell}, S. and {Crifo}, F. and {Cropper}, M. and {Crosta}, M. and {Crowley}, C. and {Dafonte}, C. and {Dapergolas}, A. and {David}, P. and {de Laverny}, P. and {De Luise}, F. and {De March}, R. and {De Ridder}, J. and {de Souza}, R. and {del Peloso}, E.~F. and {del Pozo}, E. and {Delbo}, M. and {Delgado}, A. and {Delisle}, J. -B. and {Demouchy}, C. and {Dharmawardena}, T.~E. and {Diakite}, S. and {Diener}, C. and {Distefano}, E. and {Dolding}, C. and {Enke}, H. and {Fabre}, C. and {Fabrizio}, M. and {Faigler}, S. and {Fedorets}, G. and {Fernique}, P. and {Fienga}, A. and {Figueras}, F. and {Fournier}, Y. and {Fouron}, C. and {Fragkoudi}, F. and {Gai}, M. and {Garcia-Gutierrez}, A. and {Garcia-Reinaldos}, M. and {Garc{\'\i}a-Torres}, M. and {Garofalo}, A. and {Gavel}, A. and {Gavras}, P. and {Giacobbe}, P. and {Gilmore}, G. and {Girona}, S. and {Giuffrida}, G. and {Gomel}, R. and {Gomez}, A. and {Gonz{\'a}lez-N{\'u}{\~n}ez}, J. and {Gonz{\'a}lez-Santamar{\'\i}a}, I. and {Granvik}, M. and {Guillout}, P. and {Guiraud}, J. and {Guti{\'e}rrez-S{\'a}nchez}, R. and {Guy}, L.~P. and {Hatzidimitriou}, D. and {Hauser}, M. and {Haywood}, M. and {Helmer}, A. and {Helmi}, A. and {Sarmiento}, M.~H. and {Hidalgo}, S.~L. and {H{\l}adczuk}, N. and {Holland}, G. and {Huckle}, H.~E. and {Jardine}, K. and {Jasniewicz}, G. and {Jean-Antoine Piccolo}, A. and {Jim{\'e}nez-Arranz}, {\'O}. and {Juaristi Campillo}, J. and {Julbe}, F. and {Karbevska}, L. and {Kervella}, P. and {Khanna}, S. and {Kordopatis}, G. and {Korn}, A.~J. and {K{\'o}sp{\'a}l}, {\'A}. and {Kostrzewa-Rutkowska}, Z. and {Kruszy{\'n}ska}, K. and {Kun}, M. and {Laizeau}, P. and {Lambert}, S. and {Lanza}, A.~F. and {Lasne}, Y. and {Le Campion}, J. -F. and {Lebreton}, Y. and {Lebzelter}, T. and {Leccia}, S. and {Leclerc}, N. and {Lecoeur-Taibi}, I. and {Liao}, S. and {Licata}, E.~L. and {Lindstr{\o}m}, H.~E.~P. and {Lister}, T.~A. and {Livanou}, E. and {Lobel}, A. and {Lorca}, A. and {Loup}, C. and {Madrero Pardo}, P. and {Magdaleno Romeo}, A. and {Managau}, S. and {Mann}, R.~G. and {Manteiga}, M. and {Marchant}, J.~M. and {Marconi}, M. and {Marcos}, J. and {Santos}, M.~M.~S. Marcos and {Mar{\'\i}n Pina}, D. and {Marinoni}, S. and {Marocco}, F. and {Marshall}, D.~J. and {Polo}, L. Martin and {Mart{\'\i}n-Fleitas}, J.~M. and {Marton}, G. and {Mary}, N. and {Masip}, A. and {Massari}, D. and {Mastrobuono-Battisti}, A. and {Mazeh}, T. and {Messina}, S. and {Michalik}, D. and {Millar}, N.~R. and {Mints}, A. and {Molina}, D. and {Molinaro}, R. and {Moln{\'a}r}, L. and {Monari}, G. and {Mongui{\'o}}, M. and {Montegriffo}, P. and {Montero}, A. and {Mor}, R. and {Mora}, A. and {Morbidelli}, R. and {Morel}, T. and {Morris}, D. and {Muraveva}, T. and {Murphy}, C.~P. and {Musella}, I. and {Nagy}, Z. and {Noval}, L. and {Oca{\~n}a}, F. and {Ogden}, A. and {Ordenovic}, C. and {Osinde}, J.~O. and {Pagani}, C. and {Pagano}, I. and {Palaversa}, L. and {Palicio}, P.~A. and {Pallas-Quintela}, L. and {Panahi}, A. and {Payne-Wardenaar}, S. and {Pe{\~n}alosa Esteller}, X. and {Penttil{\"a}}, A. and {Pichon}, B. and {Piersimoni}, A.~M. and {Pineau}, F. -X. and {Plachy}, E. and {Plum}, G. and {Poggio}, E. and {Pr{\v{s}}a}, A. and {Pulone}, L. and {Racero}, E. and {Ragaini}, S. and {Rainer}, M. and {Rambaux}, N. and {Ramos}, P. and {Re Fiorentin}, P. and {Regibo}, S. and {Richards}, P.~J. and {Diaz}, C. Rios and {Ripepi}, V. and {Riva}, A. and {Rix}, H. -W. and {Rixon}, G. and {Robichon}, N. and {Robin}, A.~C. and {Robin}, C. and {Roelens}, M. and {Rogues}, H.~R.~O. and {Rohrbasser}, L. and {Romero-G{\'o}mez}, M. and {Royer}, F. and {Ruz Mieres}, D. and {Rybicki}, K.~A. and {Sadowski}, G. and {S{\'a}ez N{\'u}{\~n}ez}, A. and {Sagrist{\`a} Sell{\'e}s}, A. and {Sahlmann}, J. and {Salguero}, E. and {Samaras}, N. and {Sanchez Gimenez}, V. and {Sanna}, N. and {Santove{\~n}a}, R. and {Sarasso}, M. and {Schultheis}, M. and {Sciacca}, E. and {Segol}, M. and {Segovia}, J.~C. and {S{\'e}gransan}, D. and {Semeux}, D. and {Shahaf}, S. and {Siddiqui}, H.~I. and {Siebert}, A. and {Siltala}, L. and {Silvelo}, A. and {Slezak}, E. and {Slezak}, I. and {Smart}, R.~L. and {Snaith}, O.~N. and {Solano}, E. and {Solitro}, F. and {Souami}, D. and {Souchay}, J. and {Spagna}, A. and {Spina}, L. and {Spoto}, F. and {Steele}, I.~A. and {Stephenson}, C.~A. and {S{\"u}veges}, M. and {Surdej}, J. and {Szabados}, L. and {Szegedi-Elek}, E. and {Taris}, F. and {Taylor}, M.~B. and {Teixeira}, R. and {Tolomei}, L. and {Tonello}, N. and {Torralba Elipe}, G. and {Trabucchi}, M. and {Tsounis}, A.~T. and {Turon}, C. and {Ulla}, A. and {Unger}, N. and {Vaillant}, M.~V. and {van Dillen}, E. and {van Reeven}, W. and {Vanel}, O. and {Vecchiato}, A. and {Viala}, Y. and {Vicente}, D. and {Voutsinas}, S. and {Weiler}, M. and {Wevers}, T. and {Wyrzykowski}, {\L}. and {Yoldas}, A. and {Yvard}, P. and {Zhao}, H. and {Zorec}, J. and {Zucker}, S. and {Zwitter}, T.},
        title = "{Gaia Early Data Release 3. The celestial reference frame (Gaia-CRF3)}",
      journal = {\aap},
         year = 2022,
        month = nov,
       volume = {667},
          eid = {A148},
        pages = {A148},
          doi = {10.1051/0004-6361/202243483},
archivePrefix = {arXiv},
       eprint = {2204.12574},
 primaryClass = {astro-ph.IM},
       adsurl = {https://ui.adsabs.harvard.edu/abs/2022A&A...667A.148G}
}

@ARTICLE{Husser_new_2013,
       author = {{Husser}, T. -O. and {Wende-von Berg}, S. and {Dreizler}, S. and {Homeier}, D. and {Reiners}, A. and {Barman}, T. and {Hauschildt}, P.~H.},
        title = "{A new extensive library of PHOENIX stellar atmospheres and synthetic spectra}",
      journal = {\aap},
         year = 2013,
        month = may,
       volume = {553},
          eid = {A6},
        pages = {A6},
          doi = {10.1051/0004-6361/201219058},
archivePrefix = {arXiv},
       eprint = {1303.5632},
 primaryClass = {astro-ph.SR},
       adsurl = {https://ui.adsabs.harvard.edu/abs/2013A&A...553A...6H}
}

@ARTICLE{allard_models_2012,
       author = {{Allard}, F. and {Homeier}, D. and {Freytag}, B.},
        title = "{Models of very-low-mass stars, brown dwarfs and exoplanets}",
      journal = {Philosophical Transactions of the Royal Society of London Series A},
         year = 2012,
        month = jun,
       volume = {370},
       number = {1968},
        pages = {2765-2777},
          doi = {10.1098/rsta.2011.0269},
archivePrefix = {arXiv},
       eprint = {1112.3591},
 primaryClass = {astro-ph.SR},
       adsurl = {https://ui.adsabs.harvard.edu/abs/2012RSPTA.370.2765A}
}

@ARTICLE{zarka_magnetically_2001,
       author = {{Zarka}, Philippe and {Treumann}, Rudolf A. and {Ryabov}, Boris P. and {Ryabov}, Vladimir B.},
        title = "{Magnetically-Driven Planetary Radio Emissions and Application to Extrasolar Planets}",
      journal = {\apss},
         year = 2001,
        month = jun,
       volume = {277},
        pages = {293-300},
          doi = {10.1023/A:1012221527425},
       adsurl = {https://ui.adsabs.harvard.edu/abs/2001Ap&SS.277..293Z}
}

@ARTICLE{Nichols_origin_2012,
       author = {{Nichols}, J.~D. and {Burleigh}, M.~R. and {Casewell}, S.~L. and {Cowley}, S.~W.~H. and {Wynn}, G.~A. and {Clarke}, J.~T. and {West}, A.~A.},
        title = "{Origin of Electron Cyclotron Maser Induced Radio Emissions at Ultracool Dwarfs: Magnetosphere-Ionosphere Coupling Currents}",
      journal = {\apj},
         year = 2012,
        month = nov,
       volume = {760},
       number = {1},
          eid = {59},
        pages = {59},
          doi = {10.1088/0004-637X/760/1/59},
archivePrefix = {arXiv},
       eprint = {1210.1864},
 primaryClass = {astro-ph.SR},
       adsurl = {https://ui.adsabs.harvard.edu/abs/2012ApJ...760...59N}
}

@ARTICLE{Legg_elliptic_1968,
       author = {{Legg}, M.~P.~C. and {Westfold}, K.~C.},
        title = "{Elliptic Polarization of Synchrotron Radiation}",
      journal = {\apj},
         year = 1968,
        month = nov,
       volume = {154},
        pages = {499},
          doi = {10.1086/149777},
       adsurl = {https://ui.adsabs.harvard.edu/abs/1968ApJ...154..499L}
}

@ARTICLE{Saikia_polarization_1998,
       author = {{Saikia}, D.~J. and {Holmes}, G.~F. and {Kulkarni}, A.~R. and {Salter}, C.~J. and {Garrington}, S.~T.},
        title = "{Polarization observations of the radio cores of AGN - I. A sample of quasars}",
      journal = {\mnras},
         year = 1998,
        month = aug,
       volume = {298},
       number = {3},
        pages = {877-887},
          doi = {10.1046/j.1365-8711.1998.01699.x},
       adsurl = {https://ui.adsabs.harvard.edu/abs/1998MNRAS.298..877S}
}

@ARTICLE{Callingham_radiosignatures_2024,
       author = {{Callingham}, J.~R. and {Pope}, B.~J.~S. and {Kavanagh}, R.~D. and {Bellotti}, S. and {Daley-Yates}, S. and {Damasso}, M. and {Grie{\ss}meier}, J. -M. and {G{\"u}del}, M. and {G{\"u}nther}, M. and {Kao}, M.~M. and {Klein}, B. and {Mahadevan}, S. and {Morin}, J. and {Nichols}, J.~D. and {Osten}, R.~A. and {P{\'e}rez-Torres}, M. and {Pineda}, J.~S. and {Rigney}, J. and {Saur}, J. and {Stef{\'a}nsson}, G. and {Turner}, J.~D. and {Vedantham}, H. and {Vidotto}, A.~A. and {Villadsen}, J. and {Zarka}, P.},
        title = "{Radio signatures of star-planet interactions, exoplanets and space weather}",
      journal = {Nature Astronomy},
         year = 2024,
        month = nov,
       volume = {8},
        pages = {1359-1372},
          doi = {10.1038/s41550-024-02405-6},
archivePrefix = {arXiv},
       eprint = {2409.15507},
 primaryClass = {astro-ph.EP},
       adsurl = {https://ui.adsabs.harvard.edu/abs/2024NatAs...8.1359C}
}

@INPROCEEDINGS{ramsey_early_1998,
       author = {{Ramsey}, Lawrence W. and {Adams}, M.~T. and {Barnes}, Thomas G. and {Booth}, John A. and {Cornell}, Mark E. and {Fowler}, James R. and {Gaffney}, Niall I. and {Glaspey}, John W. and {Good}, John M. and {Hill}, Gary J. and {Kelton}, Philip W. and {Krabbendam}, Victor L. and {Long}, L. and {MacQueen}, Phillip J. and {Ray}, Frank B. and {Ricklefs}, Randall L. and {Sage}, J. and {Sebring}, Thomas A. and {Spiesman}, W.~J. and {Steiner}, M.},
        title = "{Early performance and present status of the Hobby-Eberly Telescope}",
    booktitle = {Advanced Technology Optical/IR Telescopes VI},
         year = 1998,
       editor = {{Stepp}, Larry M.},
       series = {Society of Photo-Optical Instrumentation Engineers (SPIE) Conference Series},
       volume = {3352},
        month = aug,
        pages = {34-42},
          doi = {10.1117/12.319287},
       adsurl = {https://ui.adsabs.harvard.edu/abs/1998SPIE.3352...34R}
}

@ARTICLE{Hill_hetdex_2021,
       author = {{Hill}, Gary J. and {Lee}, Hanshin and {MacQueen}, Phillip J. and {Kelz}, Andreas and {Drory}, Niv and {Vattiat}, Brian L. and {Good}, John M. and {Ramsey}, Jason and {Kriel}, Herman and {Peterson}, Trent and {DePoy}, D.~L. and {Gebhardt}, Karl and {Marshall}, J.~L. and {Tuttle}, Sarah E. and {Bauer}, Svend M. and {Chonis}, Taylor S. and {Fabricius}, Maximilian H. and {Froning}, Cynthia and {H{\"a}user}, Marco and {Indahl}, Briana L. and {Jahn}, Thomas and {Landriau}, Martin and {Leck}, Ron and {Montesano}, Francesco and {Prochaska}, Travis and {Snigula}, Jan M. and {Zeimann}, Greg and {Bryant}, Randy and {Damm}, George and {Fowler}, J.~R. and {Janowiecki}, Steven and {Martin}, Jerry and {Mrozinski}, Emily and {Odewahn}, Stephen and {Rostopchin}, Sergey and {Shetrone}, Matthew and {Spencer}, Renny and {Mentuch Cooper}, Erin and {Armandroff}, Taft and {Bender}, Ralf and {Dalton}, Gavin and {Hopp}, Ulrich and {Komatsu}, Eiichiro and {Nicklas}, Harald and {Ramsey}, Lawrence W. and {Roth}, Martin M. and {Schneider}, Donald P. and {Sneden}, Chris and {Steinmetz}, Matthias},
        title = "{The HETDEX Instrumentation: Hobby-Eberly Telescope Wide-field Upgrade and VIRUS}",
      journal = {\aj},
         year = 2021,
        month = dec,
       volume = {162},
       number = {6},
          eid = {298},
        pages = {298},
          doi = {10.3847/1538-3881/ac2c02},
archivePrefix = {arXiv},
       eprint = {2110.03843},
 primaryClass = {astro-ph.IM},
       adsurl = {https://ui.adsabs.harvard.edu/abs/2021AJ....162..298H}
}

@ARTICLE{Casagrande_new_2011,
       author = {{Casagrande}, L. and {Sch{\"o}nrich}, R. and {Asplund}, M. and {Cassisi}, S. and {Ram{\'\i}rez}, I. and {Mel{\'e}ndez}, J. and {Bensby}, T. and {Feltzing}, S.},
        title = "{New constraints on the chemical evolution of the solar neighbourhood and Galactic disc(s). Improved astrophysical parameters for the Geneva-Copenhagen Survey}",
      journal = {\aap},
         year = 2011,
        month = jun,
       volume = {530},
          eid = {A138},
        pages = {A138},
          doi = {10.1051/0004-6361/201016276},
archivePrefix = {arXiv},
       eprint = {1103.4651},
 primaryClass = {astro-ph.GA},
       adsurl = {https://ui.adsabs.harvard.edu/abs/2011A&A...530A.138C}
}

@ARTICLE{Mazeh_todcor_1994,
       author = {{Mazeh}, T. and {Zucker}, S.},
        title = "{TODCOR: A Two-Dimensional Correlation Technique to Analyze Stellar Spectra in Search of Faint Companions}",
      journal = {\apss},
         year = 1994,
        month = feb,
       volume = {212},
       number = {1-2},
        pages = {349-356},
          doi = {10.1007/BF00984538},
       adsurl = {https://ui.adsabs.harvard.edu/abs/1994Ap&SS.212..349M}
}

@ARTICLE{mazeh_infrared_2002,
       author = {{Mazeh}, Tsevi and {Prato}, L. and {Simon}, M. and {Goldberg}, Elad and {Norman}, Dara and {Zucker}, Shay},
        title = "{Infrared Detection of Low-Mass Secondaries in Spectroscopic Binaries}",
      journal = {\apj},
         year = 2002,
        month = jan,
       volume = {564},
       number = {2},
        pages = {1007-1014},
          doi = {10.1086/324404},
archivePrefix = {arXiv},
       eprint = {astro-ph/0110536},
 primaryClass = {astro-ph},
       adsurl = {https://ui.adsabs.harvard.edu/abs/2002ApJ...564.1007M}
}

@ARTICLE{bender_detection_2008,
       author = {{Bender}, Chad F. and {Simon}, Michal},
        title = "{The Detection of Low-Mass Companions in Hyades Cluster Spectroscopic Binary Stars}",
      journal = {\apj},
         year = 2008,
        month = dec,
       volume = {689},
       number = {1},
        pages = {416-429},
          doi = {10.1086/592728},
archivePrefix = {arXiv},
       eprint = {0808.3393},
 primaryClass = {astro-ph},
       adsurl = {https://ui.adsabs.harvard.edu/abs/2008ApJ...689..416B}
}

@ARTICLE{bender_sdss_2012,
       author = {{Bender}, Chad F. and {Mahadevan}, Suvrath and {Deshpande}, Rohit and {Wright}, Jason T. and {Roy}, Arpita and {Terrien}, Ryan C. and {Sigurdsson}, Steinn and {Ramsey}, Lawrence W. and {Schneider}, Donald P. and {Fleming}, Scott W.},
        title = "{The SDSS-HET Survey of Kepler Eclipsing Binaries: Spectroscopic Dynamical Masses of the Kepler-16 Circumbinary Planet Hosts}",
      journal = {\apjl},
         year = 2012,
        month = jun,
       volume = {751},
       number = {2},
          eid = {L31},
        pages = {L31},
          doi = {10.1088/2041-8205/751/2/L31},
archivePrefix = {arXiv},
       eprint = {1205.0259},
 primaryClass = {astro-ph.SR},
       adsurl = {https://ui.adsabs.harvard.edu/abs/2012ApJ...751L..31B}
}

@ARTICLE{Reiners_magnetism_2022,
       author = {{Reiners}, A. and {Shulyak}, D. and {K{\"a}pyl{\"a}}, P.~J. and {Ribas}, I. and {Nagel}, E. and {Zechmeister}, M. and {Caballero}, J.~A. and {Shan}, Y. and {Fuhrmeister}, B. and {Quirrenbach}, A. and {Amado}, P.~J. and {Montes}, D. and {Jeffers}, S.~V. and {Azzaro}, M. and {B{\'e}jar}, V.~J.~S. and {Chaturvedi}, P. and {Henning}, Th. and {K{\"u}rster}, M. and {Pall{\'e}}, E.},
        title = "{Magnetism, rotation, and nonthermal emission in cool stars. Average magnetic field measurements in 292 M dwarfs}",
      journal = {\aap},
         year = 2022,
        month = jun,
       volume = {662},
          eid = {A41},
        pages = {A41},
          doi = {10.1051/0004-6361/202243251},
archivePrefix = {arXiv},
       eprint = {2204.00342},
 primaryClass = {astro-ph.SR},
       adsurl = {https://ui.adsabs.harvard.edu/abs/2022A&A...662A..41R}
}

@ARTICLE{robitaille_astropy_2013,
       author = {{Astropy Collaboration} and {Robitaille}, Thomas P. and {Tollerud}, Erik J. and {Greenfield}, Perry and {Droettboom}, Michael and {Bray}, Erik and {Aldcroft}, Tom and {Davis}, Matt and {Ginsburg}, Adam and {Price-Whelan}, Adrian M. and {Kerzendorf}, Wolfgang E. and {Conley}, Alexander and {Crighton}, Neil and {Barbary}, Kyle and {Muna}, Demitri and {Ferguson}, Henry and {Grollier}, Fr{\'e}d{\'e}ric and {Parikh}, Madhura M. and {Nair}, Prasanth H. and {Unther}, Hans M. and {Deil}, Christoph and {Woillez}, Julien and {Conseil}, Simon and {Kramer}, Roban and {Turner}, James E.~H. and {Singer}, Leo and {Fox}, Ryan and {Weaver}, Benjamin A. and {Zabalza}, Victor and {Edwards}, Zachary I. and {Azalee Bostroem}, K. and {Burke}, D.~J. and {Casey}, Andrew R. and {Crawford}, Steven M. and {Dencheva}, Nadia and {Ely}, Justin and {Jenness}, Tim and {Labrie}, Kathleen and {Lim}, Pey Lian and {Pierfederici}, Francesco and {Pontzen}, Andrew and {Ptak}, Andy and {Refsdal}, Brian and {Servillat}, Mathieu and {Streicher}, Ole},
        title = "{Astropy: A community Python package for astronomy}",
      journal = {\aap},
         year = 2013,
        month = oct,
       volume = {558},
          eid = {A33},
        pages = {A33},
          doi = {10.1051/0004-6361/201322068},
archivePrefix = {arXiv},
       eprint = {1307.6212},
 primaryClass = {astro-ph.IM},
       adsurl = {https://ui.adsabs.harvard.edu/abs/2013A&A...558A..33A}
}

@ARTICLE{astropy_collaboration_astropy_2018,
       author = {{Astropy Collaboration} and {Price-Whelan}, A.~M. and {Sip{\H{o}}cz}, B.~M. and {G{\"u}nther}, H.~M. and {Lim}, P.~L. and {Crawford}, S.~M. and {Conseil}, S. and {Shupe}, D.~L. and {Craig}, M.~W. and {Dencheva}, N. and {Ginsburg}, A. and {VanderPlas}, J.~T. and {Bradley}, L.~D. and {P{\'e}rez-Su{\'a}rez}, D. and {de Val-Borro}, M. and {Aldcroft}, T.~L. and {Cruz}, K.~L. and {Robitaille}, T.~P. and {Tollerud}, E.~J. and {Ardelean}, C. and {Babej}, T. and {Bach}, Y.~P. and {Bachetti}, M. and {Bakanov}, A.~V. and {Bamford}, S.~P. and {Barentsen}, G. and {Barmby}, P. and {Baumbach}, A. and {Berry}, K.~L. and {Biscani}, F. and {Boquien}, M. and {Bostroem}, K.~A. and {Bouma}, L.~G. and {Brammer}, G.~B. and {Bray}, E.~M. and {Breytenbach}, H. and {Buddelmeijer}, H. and {Burke}, D.~J. and {Calderone}, G. and {Cano Rodr{\'\i}guez}, J.~L. and {Cara}, M. and {Cardoso}, J.~V.~M. and {Cheedella}, S. and {Copin}, Y. and {Corrales}, L. and {Crichton}, D. and {D'Avella}, D. and {Deil}, C. and {Depagne}, {\'E}. and {Dietrich}, J.~P. and {Donath}, A. and {Droettboom}, M. and {Earl}, N. and {Erben}, T. and {Fabbro}, S. and {Ferreira}, L.~A. and {Finethy}, T. and {Fox}, R.~T. and {Garrison}, L.~H. and {Gibbons}, S.~L.~J. and {Goldstein}, D.~A. and {Gommers}, R. and {Greco}, J.~P. and {Greenfield}, P. and {Groener}, A.~M. and {Grollier}, F. and {Hagen}, A. and {Hirst}, P. and {Homeier}, D. and {Horton}, A.~J. and {Hosseinzadeh}, G. and {Hu}, L. and {Hunkeler}, J.~S. and {Ivezi{\'c}}, {\v{Z}}. and {Jain}, A. and {Jenness}, T. and {Kanarek}, G. and {Kendrew}, S. and {Kern}, N.~S. and {Kerzendorf}, W.~E. and {Khvalko}, A. and {King}, J. and {Kirkby}, D. and {Kulkarni}, A.~M. and {Kumar}, A. and {Lee}, A. and {Lenz}, D. and {Littlefair}, S.~P. and {Ma}, Z. and {Macleod}, D.~M. and {Mastropietro}, M. and {McCully}, C. and {Montagnac}, S. and {Morris}, B.~M. and {Mueller}, M. and {Mumford}, S.~J. and {Muna}, D. and {Murphy}, N.~A. and {Nelson}, S. and {Nguyen}, G.~H. and {Ninan}, J.~P. and {N{\"o}the}, M. and {Ogaz}, S. and {Oh}, S. and {Parejko}, J.~K. and {Parley}, N. and {Pascual}, S. and {Patil}, R. and {Patil}, A.~A. and {Plunkett}, A.~L. and {Prochaska}, J.~X. and {Rastogi}, T. and {Reddy Janga}, V. and {Sabater}, J. and {Sakurikar}, P. and {Seifert}, M. and {Sherbert}, L.~E. and {Sherwood-Taylor}, H. and {Shih}, A.~Y. and {Sick}, J. and {Silbiger}, M.~T. and {Singanamalla}, S. and {Singer}, L.~P. and {Sladen}, P.~H. and {Sooley}, K.~A. and {Sornarajah}, S. and {Streicher}, O. and {Teuben}, P. and {Thomas}, S.~W. and {Tremblay}, G.~R. and {Turner}, J.~E.~H. and {Terr{\'o}n}, V. and {van Kerkwijk}, M.~H. and {de la Vega}, A. and {Watkins}, L.~L. and {Weaver}, B.~A. and {Whitmore}, J.~B. and {Woillez}, J. and {Zabalza}, V. and {Astropy Contributors}},
        title = "{The Astropy Project: Building an Open-science Project and Status of the v2.0 Core Package}",
      journal = {\aj},
         year = 2018,
        month = sep,
       volume = {156},
       number = {3},
          eid = {123},
        pages = {123},
          doi = {10.3847/1538-3881/aabc4f},
archivePrefix = {arXiv},
       eprint = {1801.02634},
 primaryClass = {astro-ph.IM},
       adsurl = {https://ui.adsabs.harvard.edu/abs/2018AJ....156..123A}
}

@ARTICLE{astropy_2022,
       author = {{Astropy Collaboration} and {Price-Whelan}, Adrian M. and {Lim}, Pey Lian and {Earl}, Nicholas and {Starkman}, Nathaniel and {Bradley}, Larry and {Shupe}, David L. and {Patil}, Aarya A. and {Corrales}, Lia and {Brasseur}, C.~E. and {N{\"o}the}, Maximilian and {Donath}, Axel and {Tollerud}, Erik and {Morris}, Brett M. and {Ginsburg}, Adam and {Vaher}, Eero and {Weaver}, Benjamin A. and {Tocknell}, James and {Jamieson}, William and {van Kerkwijk}, Marten H. and {Robitaille}, Thomas P. and {Merry}, Bruce and {Bachetti}, Matteo and {G{\"u}nther}, H. Moritz and {Aldcroft}, Thomas L. and {Alvarado-Montes}, Jaime A. and {Archibald}, Anne M. and {B{\'o}di}, Attila and {Bapat}, Shreyas and {Barentsen}, Geert and {Baz{\'a}n}, Juanjo and {Biswas}, Manish and {Boquien}, M{\'e}d{\'e}ric and {Burke}, D.~J. and {Cara}, Daria and {Cara}, Mihai and {Conroy}, Kyle E. and {Conseil}, Simon and {Craig}, Matthew W. and {Cross}, Robert M. and {Cruz}, Kelle L. and {D'Eugenio}, Francesco and {Dencheva}, Nadia and {Devillepoix}, Hadrien A.~R. and {Dietrich}, J{\"o}rg P. and {Eigenbrot}, Arthur Davis and {Erben}, Thomas and {Ferreira}, Leonardo and {Foreman-Mackey}, Daniel and {Fox}, Ryan and {Freij}, Nabil and {Garg}, Suyog and {Geda}, Robel and {Glattly}, Lauren and {Gondhalekar}, Yash and {Gordon}, Karl D. and {Grant}, David and {Greenfield}, Perry and {Groener}, Austen M. and {Guest}, Steve and {Gurovich}, Sebastian and {Handberg}, Rasmus and {Hart}, Akeem and {Hatfield-Dodds}, Zac and {Homeier}, Derek and {Hosseinzadeh}, Griffin and {Jenness}, Tim and {Jones}, Craig K. and {Joseph}, Prajwel and {Kalmbach}, J. Bryce and {Karamehmetoglu}, Emir and {Ka{\l}uszy{\'n}ski}, Miko{\l}aj and {Kelley}, Michael S.~P. and {Kern}, Nicholas and {Kerzendorf}, Wolfgang E. and {Koch}, Eric W. and {Kulumani}, Shankar and {Lee}, Antony and {Ly}, Chun and {Ma}, Zhiyuan and {MacBride}, Conor and {Maljaars}, Jakob M. and {Muna}, Demitri and {Murphy}, N.~A. and {Norman}, Henrik and {O'Steen}, Richard and {Oman}, Kyle A. and {Pacifici}, Camilla and {Pascual}, Sergio and {Pascual-Granado}, J. and {Patil}, Rohit R. and {Perren}, Gabriel I. and {Pickering}, Timothy E. and {Rastogi}, Tanuj and {Roulston}, Benjamin R. and {Ryan}, Daniel F. and {Rykoff}, Eli S. and {Sabater}, Jose and {Sakurikar}, Parikshit and {Salgado}, Jes{\'u}s and {Sanghi}, Aniket and {Saunders}, Nicholas and {Savchenko}, Volodymyr and {Schwardt}, Ludwig and {Seifert-Eckert}, Michael and {Shih}, Albert Y. and {Jain}, Anany Shrey and {Shukla}, Gyanendra and {Sick}, Jonathan and {Simpson}, Chris and {Singanamalla}, Sudheesh and {Singer}, Leo P. and {Singhal}, Jaladh and {Sinha}, Manodeep and {Sip{\H{o}}cz}, Brigitta M. and {Spitler}, Lee R. and {Stansby}, David and {Streicher}, Ole and {{\v{S}}umak}, Jani and {Swinbank}, John D. and {Taranu}, Dan S. and {Tewary}, Nikita and {Tremblay}, Grant R. and {de Val-Borro}, Miguel and {Van Kooten}, Samuel J. and {Vasovi{\'c}}, Zlatan and {Verma}, Shresth and {de Miranda Cardoso}, Jos{\'e} Vin{\'\i}cius and {Williams}, Peter K.~G. and {Wilson}, Tom J. and {Winkel}, Benjamin and {Wood-Vasey}, W.~M. and {Xue}, Rui and {Yoachim}, Peter and {Zhang}, Chen and {Zonca}, Andrea and {Astropy Project Contributors}},
        title = "{The Astropy Project: Sustaining and Growing a Community-oriented Open-source Project and the Latest Major Release (v5.0) of the Core Package}",
      journal = {\apj},
         year = 2022,
        month = aug,
       volume = {935},
       number = {2},
          eid = {167},
        pages = {167},
          doi = {10.3847/1538-4357/ac7c74},
archivePrefix = {arXiv},
       eprint = {2206.14220},
 primaryClass = {astro-ph.IM},
       adsurl = {https://ui.adsabs.harvard.edu/abs/2022ApJ...935..167A}
}

@ARTICLE{perez_ipython_2007,
       author = {{Perez}, Fernando and {Granger}, Brian E.},
        title = "{IPython: A System for Interactive Scientific Computing}",
      journal = {Computing in Science and Engineering},
         year = 2007,
        month = jan,
       volume = {9},
       number = {3},
        pages = {21-29},
          doi = {10.1109/MCSE.2007.53},
       adsurl = {https://ui.adsabs.harvard.edu/abs/2007CSE.....9c..21P}
}

@Article{hunter_matplotlib_2007,
  Author    = {Hunter, J. D.},
  Title     = {Matplotlib: A 2D graphics environment},
  Journal   = {Computing in Science \& Engineering},
  Volume    = {9},
  Number    = {3},
  Pages     = {90--95},
  publisher = {IEEE COMPUTER SOC},
  doi       = {10.1109/MCSE.2007.55},
  year      = 2007
}

@Article{harris_numpy_2020,
 title         = {Array programming with {NumPy}},
 author        = {Charles R. Harris and K. Jarrod Millman and St{\'{e}}fan J.
                 van der Walt and Ralf Gommers and Pauli Virtanen and David
                 Cournapeau and Eric Wieser and Julian Taylor and Sebastian
                 Berg and Nathaniel J. Smith and Robert Kern and Matti Picus
                 and Stephan Hoyer and Marten H. van Kerkwijk and Matthew
                 Brett and Allan Haldane and Jaime Fern{\'{a}}ndez del
                 R{\'{i}}o and Mark Wiebe and Pearu Peterson and Pierre
                 G{\'{e}}rard-Marchant and Kevin Sheppard and Tyler Reddy and
                 Warren Weckesser and Hameer Abbasi and Christoph Gohlke and
                 Travis E. Oliphant},
 year          = {2020},
 month         = sep,
 journal       = {Nature},
 volume        = {585},
 number        = {7825},
 pages         = {357--362},
 doi           = {10.1038/s41586-020-2649-2},
 publisher     = {Springer Science and Business Media {LLC}},
 url           = {https://doi.org/10.1038/s41586-020-2649-2}
}

@InProceedings{mckinney_data_2010,
  author    = { {W}es {M}c{K}inney },
  title     = { {D}ata {S}tructures for {S}tatistical {C}omputing in {P}ython },
  booktitle = { {P}roceedings of the 9th {P}ython in {S}cience {C}onference },
  pages     = { 56 - 61 },
  year      = { 2010 },
  editor    = { {S}t\'efan van der {W}alt and {J}arrod {M}illman },
  doi       = { 10.25080/Majora-92bf1922-00a }
}

@software{reback2020pandas,
    author       = {The pandas development team},
    title        = {pandas-dev/pandas: Pandas},
    month        = feb,
    year         = 2020,
    publisher    = {Zenodo},
    version      = {latest},
    doi          = {10.5281/zenodo.3509134},
    url          = {https://doi.org/10.5281/zenodo.3509134}
}

@ARTICLE{virtanen_scipy_2020,
  author  = {Virtanen, Pauli and Gommers, Ralf and Oliphant, Travis E. and
            Haberland, Matt and Reddy, Tyler and Cournapeau, David and
            Burovski, Evgeni and Peterson, Pearu and Weckesser, Warren and
            Bright, Jonathan and {van der Walt}, St{\'e}fan J. and
            Brett, Matthew and Wilson, Joshua and Millman, K. Jarrod and
            Mayorov, Nikolay and Nelson, Andrew R. J. and Jones, Eric and
            Kern, Robert and Larson, Eric and Carey, C J and
            Polat, {\.I}lhan and Feng, Yu and Moore, Eric W. and
            {VanderPlas}, Jake and Laxalde, Denis and Perktold, Josef and
            Cimrman, Robert and Henriksen, Ian and Quintero, E. A. and
            Harris, Charles R. and Archibald, Anne M. and
            Ribeiro, Ant{\^o}nio H. and Pedregosa, Fabian and
            {van Mulbregt}, Paul and {SciPy 1.0 Contributors}},
  title   = {{{SciPy} 1.0: Fundamental Algorithms for Scientific
            Computing in Python}},
  journal = {Nature Methods},
  year    = {2020},
  volume  = {17},
  pages   = {261--272},
  adsurl  = {https://rdcu.be/b08Wh},
  doi     = {10.1038/s41592-019-0686-2},
}

@ARTICLE{gudel_radio_1992,
       author = {{Gudel}, M.},
        title = "{Radio and X-ray emission from main-sequence K stars.}",
      journal = {\aap},
         year = 1992,
        month = oct,
       volume = {264},
        pages = {L31-L34},
       adsurl = {https://ui.adsabs.harvard.edu/abs/1992A&A...264L..31G}
}

@ARTICLE{Gudel_tight_1993,
       author = {{Gudel}, Manuel and {Schmitt}, Juergen H.~M.~M. and {Bookbinder}, Jay A. and {Fleming}, Thomas A.},
        title = "{A Tight Correlation between Radio and X-Ray Luminosities of M Dwarfs}",
      journal = {\apj},
         year = 1993,
        month = sep,
       volume = {415},
        pages = {236},
          doi = {10.1086/173158},
       adsurl = {https://ui.adsabs.harvard.edu/abs/1993ApJ...415..236G}
}

@ARTICLE{benz_xray_1994,
       author = {{Benz}, A.~O. and {Guedel}, M.},
        title = "{X-ray/microwave ratio of flares and coronae}",
      journal = {\aap},
         year = 1994,
        month = may,
       volume = {285},
        pages = {621-630},
       adsurl = {https://ui.adsabs.harvard.edu/abs/1994A&A...285..621B}
}

@ARTICLE{Henden_apass_2014,
       author = {{Henden}, A. and {Munari}, U.},
        title = "{The APASS all-sky, multi-epoch BVgri photometric survey}",
      journal = {Contributions of the Astronomical Observatory Skalnate Pleso},
         year = 2014,
        month = mar,
       volume = {43},
       number = {3},
        pages = {518-522},
       adsurl = {https://ui.adsabs.harvard.edu/abs/2014CoSka..43..518H}
}

@ARTICLE{bianchi_galex_2011,
       author = {{Bianchi}, L. and {Herald}, J. and {Efremova}, B. and {Girardi}, L. and {Zabot}, A. and {Marigo}, P. and {Conti}, A. and {Shiao}, B.},
        title = "{GALEX catalogs of UV sources: statistical properties and sample science applications: hot white dwarfs in the Milky Way}",
      journal = {\apss},
         year = 2011,
        month = sep,
       volume = {335},
       number = {1},
        pages = {161-169},
          doi = {10.1007/s10509-010-0581-x},
       adsurl = {https://ui.adsabs.harvard.edu/abs/2011Ap&SS.335..161B}
}

@ARTICLE{Speagle_dynesty_2020,
       author = {{Speagle}, Joshua S.},
        title = "{DYNESTY: a dynamic nested sampling package for estimating Bayesian posteriors and evidences}",
      journal = {\mnras},
         year = 2020,
        month = apr,
       volume = {493},
       number = {3},
        pages = {3132-3158},
          doi = {10.1093/mnras/staa278},
archivePrefix = {arXiv},
       eprint = {1904.02180},
 primaryClass = {astro-ph.IM},
       adsurl = {https://ui.adsabs.harvard.edu/abs/2020MNRAS.493.3132S}
}

@MISC{isochrones,
       author = {{Morton}, Timothy D.},
        title = "{isochrones: Stellar model grid package}",
         year = "2015",
        month = "Mar",
          eid = {ascl:1503.010},
        pages = {ascl:1503.010},
archivePrefix = {ascl},
       eprint = {1503.010},
       adsurl = {https://ui.adsabs.harvard.edu/abs/2015ascl.soft03010M}
}

@ARTICLE{MIST,
       author = {{Dotter}, Aaron},
        title = "{MESA Isochrones and Stellar Tracks (MIST) 0: Methods for the Construction of Stellar Isochrones}",
      journal = {\apjs},
         year = "2016",
        month = "Jan",
       volume = {222},
       number = {1},
          eid = {8},
        pages = {8},
          doi = {10.3847/0067-0049/222/1/8},
archivePrefix = {arXiv},
       eprint = {1601.05144},
 primaryClass = {astro-ph.SR},
       adsurl = {https://ui.adsabs.harvard.edu/abs/2016ApJS..222....8D}
}

@ARTICLE{Men_highly_2025,
       author = {{Men}, Yunpeng and {McSweeney}, Sam and {Hurley-Walker}, Natasha and {Barr}, Ewan and {Stappers}, Ben},
        title = "{A highly magnetized long-period radio transient exhibiting unusual emission features}",
      journal = {Science Advances},
         year = 2025,
        month = jan,
       volume = {11},
       number = {3},
          eid = {eadp6351},
        pages = {eadp6351},
          doi = {10.1126/sciadv.adp6351},
archivePrefix = {arXiv},
       eprint = {2501.10528},
 primaryClass = {astro-ph.HE},
       adsurl = {https://ui.adsabs.harvard.edu/abs/2025SciA...11P6351M}
}
\bibliographystyle{aasjournal}



\end{document}